\documentclass[article]{IEEEtran}

\usepackage[pass]{geometry}

\usepackage{t1enc}
\usepackage[utf8]{inputenc}

\usepackage{cite}

\usepackage{etoolbox}

\usepackage{textcomp}
\usepackage{wrapfig}
\usepackage{lscape}
\usepackage{verbatim}
\usepackage{url}
\usepackage{color}
\usepackage{amsmath}
\usepackage{graphicx}
\usepackage{multirow}
\usepackage{booktabs}
\usepackage{listings}
\usepackage{ifthen}

\usepackage[noend]{algpseudocode}
\usepackage{algorithm}

\usepackage{subcaption}

\usepackage{xcolor}
\usepackage[hidelinks]{hyperref}

\usepackage{tikz,pgfplots}

\usepackage{placeins}

\newcommand{\ceil}[1]{\left\lceil #1\right\rceil}
\newcommand{\floor}[1]{\left\lfloor #1\right\rfloor}
\newcommand{\abs}[1]{\left| #1\right|}

\newcommand{\set}[1]{\left\{ #1\right\}}

\newcommand{\realrange}[2]{\left[#1, #2\right]}

\newcommand{\unitrange}[2]{\realrange{0}{1}}

\newcommand{\Oh}[1]{\mathcal{O}\!\left( #1\right)}

\newcommand{\Th}[1]{\Theta\!\left( #1\right)}
\newcommand{\Om}[1]{\Omega\left(#1\right)}

\newcommand{\llabel}[1]{\label{\labelprefix:#1}}
\newcommand{\labelprefix}{} 

\newcommand{\discussionsize}{\small}

\newcommand{\frage}[1]{{[\bf #1]}\marginpar{$\bigotimes$}}

\newcommand{\punkt}{\enspace .}

\newenvironment{code}{\noindent\normalsize
\begin{tabbing}%
\hspace{2em}\=\hspace{2em}\=\hspace{2em}\=\hspace{2em}\=\hspace{2em}\=%
\hspace{2em}\=\hspace{2em}\=\hspace{2em}\=\hspace{2em}\=\hspace{2em}\=%
\kill}{\end{tabbing}}

\newcommand{\labelcommand}{}
\newcommand{\captiontext}{}
\newsavebox{\codeparam}
\newcounter{lineNumber}
\newenvironment{disscodepos}[3]{%
\renewcommand{\labelcommand}{#2}%
\renewcommand{\captiontext}{#3}%
\sbox{\codeparam}{\parbox{\textwidth}{#3}}%
\begin{figure}[#1]\begin{center}\begin{code}\setcounter{lineNumber}{1}}{%
\end{code}\end{center}\caption{\llabel{\labelcommand}\captiontext}\end{figure}}

{\end{disscodepos}}

\newdimen\endofsize\endofsize=0.5em

\newcommand{\agather}{GatherM\xspace}
\newcommand{\aallgather}{AllGatherM\xspace}
\newcommand{\aagather}{(All-)GatherM\xspace}
\newcommand{\afis}{RFIS\xspace}
\newcommand{\aquick}{RQuick\xspace}
\newcommand{\aams}{RAMS\xspace}
\newcommand{\asamplesort}{SSort\xspace}
\newcommand{\ahyk}{HykSort\xspace}
\newcommand{\abitonic}{Bitonic\xspace}
\newcommand{\aminisort}{Minisort\xspace}

\newcommand{\antbquick}{NTB-Quick\xspace}
\newcommand{\antbams}{NTB-AMS\xspace}
\newcommand{\andaams}{NDMA-AMS\xspace}
\newcommand{\andsams}{NS-SSort\xspace}
\newcommand{\afir}{FIR\xspace}

\newcommand{\adma}{DMA\xspace}

\newcommand{\bucketsorted}{BucketSorted\xspace}
\newcommand{\deterdupl}{DeterDupl\xspace}
\newcommand{\gaussian}{Gaussian\xspace}
\newcommand{\ggroup}{g-Group\xspace}
\newcommand{\randdupl}{RandDupl\xspace}
\newcommand{\staggered}{Staggered\xspace}
\newcommand{\uniform}{Uniform\xspace}
\newcommand{\zero}{Zero\xspace}
\newcommand{\reverse}{Reverse\xspace}
\newcommand{\alltoone}{AllToOne\xspace}
\newcommand{\mirrored}{Mirrored\xspace}
\newcommand{\Is}{\mbox{\rm := }}

\renewcommand{\Oh}[1]{\ensuremath{\mathcal{O}(#1)}}
\newcommand{\OhL}[1]{\ensuremath{\mathcal{O}\!\left(#1\right)}}

\renewcommand{\Om}[1]{\ensuremath{\Omega(#1)}}

\newcommand{\OmL}[1]{\ensuremath{\Omega\!\left(#1\right)}}

\renewcommand{\Th}[1]{\Theta(#1)}

\newcommand{\ceiling}[1]{\lceil #1\rceil}
\newcommand{\flooring}[1]{\lfloor #1\rfloor}

\newcommand{\postponed}[1]{}

\usepackage{balance}

\usepackage{etex}
\usepackage{mathtools}

\usepackage{xspace}

\usepackage{siunitx}

\newtoggle{extended}
\togglefalse{extended}

\newtheorem{conjecture}{Conjecture}

\renewcommand{\frage}[1]{}

\newcommand{\todo}[1]{}

\newcommand{\anmerkung}[1]{}

\begin{document}

\title{Robust Massively Parallel Sorting}

\author{Michael Axtmann, Peter Sanders\\ 
	\textit{Karlsruhe Institute of Technology},
	\textit{Karlsruhe, Germany} \\
	\textit{\normalsize\{\url{michael.axtmann, sanders}\}\url{@kit.edu}}}

\maketitle

\begin{abstract}
We investigate distributed memory parallel sorting algorithms that scale to the largest available machines and are robust with respect to input size and distribution of the input elements.
The main outcome is that four sorting algorithms cover the entire range of possible input sizes.
For three algorithms we devise new low overhead mechanisms to make them robust with respect to duplicate keys and skewed input distributions.
One of these, designed for medium sized inputs, is a new variant of quicksort with fast high-quality pivot selection.

At the same time asymptotic analysis provides performance guarantees and guides the selection and configuration of the algorithms.
We validate these hypotheses using extensive experiments on 7 algorithms, 10 input distributions, up to 262\,144 cores, and varying input sizes over 9 orders of magnitude. 
For ``difficult'' input distributions, our algorithms are the only ones that work at all. 
For all but the largest input sizes, we are the first to perform experiments on such large machines at all and our algorithms significantly outperform the ones one would conventionally have considered.

\end{abstract}

\section{Introduction}
\frage{todo: cite Ramachandran}
Sorting is one of the most fundamental and widely used non-numerical algorithms.
It is used to build index data structures, to establish invariants for further processing, to bring together ``similar'' data, and in many other applications. This wide variety of applications means that we need fast sorting algorithms for a wide spectrum of inputs with respect to data type, input size, and distribution of keys.
The motivation for this paper is that parallel algorithms currently do not cover this spectrum of possible inputs for very large parallel machines.
Although hundreds of papers on parallel sorting have been published, there is only a small number of practical studies that consider the largest machines with many thousands of processors (PEs).
The few studies known to us mostly concentrate on very large random inputs. Most of these algorithms become slow (or simply crash) when applied to worst case inputs where the location of the input data is correlated with key values (skewed inputs) or which may have large numbers of duplicated keys.

Even for random inputs, these algorithms are slow on small inputs, where their running time is dominated by a large number of message startup overheads. Note that sorting small inputs becomes important for the performance of parallel applications when it is needed in some frequently repeated coordination step between the PEs. For example, many applications perform load (re)balancing by mapping objects to space filling curves and sorting them with respect to this ordering \cite{bader2012space}. The scalability of the sorting algorithm may then become the limiting factor for the number of time steps we can do per second. Note that in this case it is of secondary importance whether the sorting algorithm itself is efficient compared to a sequential one -- what matters is that it is faster than the local computations between rebalancing steps. Siebert and Wolf \cite{SieWol11} give another extreme example, pointing out that the {\tt MPI\_Comm\_Split} operation requires sorting with exactly one input per PE. 

The subject of this paper is to systematically explore the design space of parallel algorithms for massively parallel machines and propose robust algorithms for the entire spectrum of possible inputs. As a first design decision, we restrict ourselves to comparison-based algorithms because they work for arbitrary data types and are usually less dependent on the distribution of key values than non-comparison based algorithms like radix sort. The disadvantage of more local work for comparison-based algorithms is often irrelevant for the largest machines since communication overhead is the main bottleneck. For similar reasons, we abstain from low-level tuning like exploiting SIMD-instructions or using hybrid codes with node-local shared memory algorithms. 

Let $n$ denote the input size and $p$ the number of PEs. We assume that each PE has a local input of size $\Oh{\frac{n}{p}}$, i.e, the input is ``reasonably'' balanced. We consider three major issues with respect to the robustness of a massively parallel sorting algorithm: Its scalability, i.e., its running time as a function of $n$ and $p$, how the algorithm behaves with respect to skewed input distributions and how it handles repeatedly occurring keys. 

\emph{Our Contributions.}
A methodological contribution is that
we address the scalability issue by reconciling theory and practice (to some extent). 
Classical theoretical algorithms concentrate on the issue how fast sorting algorithms can be made for small inputs.
For example, Cole's celebrated parallel mergesort \cite{Col88} runs in time $\Oh{\log p}$ for $n=p$ on a PRAM.
However, this result and the PRAM machine model are considered quite impractical. We show that simpler algorithms with polylogarithmic running time are actually practical for small $n/p$.
Using asymptotic analysis in a more realistic yet simple complexity model that distinguishes between startup latencies and communication bandwidth, we get a consistent ranking of a large number of algorithms with respect to their performance over the entire range of values for $n/p$.
We propose three robust sorting algorithms for different input sizes
which can be used to span the entire parameter space.
The first sorts while data is routed to a single PE,
the second is a simple yet fast work-inefficient algorithm with logarithmic latency (\afis), the third is an efficient variant of parallel Quicksort with latency $\Oh{\log^2 p}$ (\aquick) and the fourth has latency $\Oh{p^{1/d}}$ when we allow data to be moved $d$ times
(\aams).

\afis and \aams improve our own results from a mostly theoretical paper~\cite{Axtmann15} that includes a nonrobust proof of concept implementation and experiments on up to $2^{15}$ PEs.
However, these experiments only used uniform inputs and had highly fluctuating running times.
We also concentrated on rather large inputs (\afis is used only as a nonrobust subroutine for ranking samples).
Our contribution to \afis and \aams are robust implementations and extensive experiments with various input distributions on up  to $2^{18}$ PEs. 
We use the input distributions from Helman et al.\cite{Helman98}, who pioneered robustness experiments on parallel sorting algorithms.
Our quicksort algorithm is new, incorporating a fast and accurate splitter selection algorithm and low overhead measures to handle skew and repeated keys.
We conjecture that the expected asymptotic running time for worst case inputs (skewed but with unique keys) is the same as for best case inputs where local data is split perfectly in each step.
This algorithm closes the gap between very small and large inputs where \aams is still inefficient but the local work of \afis begins to dominate its startup latencies.
Finally, we compare our algorithms directly with several state of the art implementations considered in a recent study \cite{Sundar13}.
It turns out that sorting while gathering data on a single PE ``sorts'' very sparse inputs the fastest.
\afis is the fastest algorithm on sparse and very small inputs.
Our new quicksort outperforms its competitors on small inputs, in the range of $2^3$ to $2^{14}$ elements per core, robustly for all input instances.
HykSort \cite{Sundar13} is slightly faster than \aams for large random inputs but less robust.
Classical sample sort and related algorithms that communicate the data only once are very slow even for rather large $n/p$ \cite{BleEtAl91,VarEtAl91,GerVal94, Helman98,SSP07}.
Classical bitonic sorting \cite{Batcher68,johnsson84} is somewhat competitive only for a rather narrow range of input sizes.

\emph{Paper Overview.}
We give preliminary definitions in Section~\ref{s:prelim} and describe building blocks for random shuffling and median estimation in Section~\ref{s:building}. Section~\ref{s:related} compares eleven parallel sorting algorithms analytically and qualitatively. Our main new algorithms are described in more detail in Sections~\ref{s:fastInefficient ams}--\ref{s:quicksort}. Section~\ref{s:experiments} describes an extensive experimental evaluation.

\section{Preliminaries}\label{s:prelim}
\label{s:hypercube architecture}
\frage{ps:demoted subsections to paragraphs}

The input of sorting algorithms are $n$ elements with $\Oh{n/p}$ elements on each PE. The output must be globally sorted, i.e., each PE has elements with consecutive ranks. We also want $\Oh{n/p}$ output elements on each PE. Sometimes we more concretely consider perfectly balanced inputs and an output with at most $(1 + \epsilon)n/p$ elements per PE for some small positive constant $\epsilon$.

A naive approach to generating unique keys appends unique identifiers to the elements.
Then, lexicographic ordering makes the keys unique.
However, this approach makes sorting considerably more expensive due to higher communication volume and more expensive comparisons.

\emph{Model of Computation.}
A common abstraction of communication in supercomputers is the (symmetric) single-ported message passing model. Algorithms in this model are often implemented using the MPI, a standardized message passing interface~\cite{GroEtAl95},
that supports point-to-point message exchange as well as a wide variety of collective operations like broadcast, reduction, prefix sum,~etc.
It takes time $\alpha + l\beta$ to send a message of size $l$ machine words.
The parameter $\alpha$ defines the startup overhead of the communication, while $\beta$ defines the  time to communicate a machine word.
Appendix~\ref{app:model of computation} gives further details of the model of computation.

\emph{Hypercube Algorithms.}
\begin{algorithm}
\begin{algorithmic}
  \caption{Hypercube algorithm design pattern}\label{algo:hypercube}
  \State Computation on PE $i$
  \For{$j\Is d - 1$ {\bf downto} 0} \Comment{or $0..d-1$}
    \State send some message $m$ to PE $i\oplus 2^j$
    \State receive message $m'$ from PE $i\oplus 2^j$
    \State perform some local computation using $m$ and $m'$
  \EndFor
\end{algorithmic}
\end{algorithm}
A hypercube network of dimension $d$ consists of $p=2^d$ PEs numbered $\set{0,\ldots,p-1}$.
Two nodes $a$ and $b$ are connected along dimension $i$ if $a=b\oplus 2^i$.
For this paper, hypercubes are not primarily important as an actual network architecture. Rather, we extensively use communication in a conceptual hypercube as a design pattern for algorithms.
More specifically the hypercube algorithms shown in Algorithm~\ref{algo:hypercube} iterate through the dimension of the hypercube. 
Depending on how this template is instantiated, one achieves a large spectrum of global effects. To describe and understand hypercube algorithms, we need the concept of a subcube. A $j$-dimensional subcube consists of those PEs whose numbers have the same bits $j..d-1$ in their binary representation.
Basic operations such as \emph{all-reduce}, \emph{all-gather}, and routing data for random start or destination nodes~\cite{Lei92} need only $\Oh{\alpha \log p}$ startup overheads overall.
We call the hypercube algorithm which places all elements on all PEs in sorted order \emph{all-gather-merge}. This operation runs in time $\Oh{\beta p |a| + \alpha\log p}$.
Appendix~\ref{app:hypercube architecture} describes these operations in more detail.

\section{Building Blocks}\label{s:building}

\subsection{Randomized Shuffling on Hypercubes}\label{s:randomShuffle}

It is folklore (e.g. \cite{sanders97}) that data skew can be removed by shuffling the input data randomly. Helman et al.~\cite{Helman98} implemented this for the sample sort algorithm by directly sending each element to a random destination. Note that this incurs an overhead of about $\alpha p+\beta n/p$.
In Appendix~\ref{app:randomShuffle} we propose a technique for small inputs which runs in
$\Oh{(\alpha+\beta \frac{n}{p})\log p}$.

\subsection{Approximate Median Selection with a Single Reduction}
\label{s:approxRankSelection}
Siebert and Wolf~\cite{SieWol11} consider the case where $n=p$ and propose to select splitters  for parallel quicksort using a ternary tree
whose leaves are the input elements.
At each internal node, the median of the children elements is passed upward. Dean et al.~\cite{dean2014} show that for randomly permuted inputs and a balanced tree this is a good approximation of the median, i.e., with high probability, one gets rank $n/2(1\pm 2n^{-0.37})$.
Siebert and Wolf~\cite{SieWol11} do not permute randomly.
Even when $p$ is a power of two, their method does not produce a completely balanced tree so that their implementation remains a heuristic.

We fix these restrictions by using a binary tree and working with randomly permuted inputs:
Consider a tuning parameter $k$ (even).
Assume that the local data is a sorted sequence $a[1..m]$ and that $m$ is even. 
Each PE is a leaf of the binary tree and forwards $a[m/2-k/2+1..m/2+k/2]$
up the tree -- a local approximation of the elements closest to the median. Undefined array entries to the left are treated like a very small key and undefined entries to the right as very large keys. If $m$ is odd, we flip a coin whether $a[\floor{m/2}-k/2+1..\floor{m/2}+k/2]$ or $a[\ceil{m/2}-k/2+1..\ceil{m/2}+k/2]$ is forwarded.
Internal nodes merge the received sequences and use the result 
as their sequence $a$ analogously to the leaves.
At the root, we flip a coin whether $a[k/2]$ or $a[k/2+1]$ is returned. Note that in most MPI~implementations, this algorithm can be implemented by defining an appropriate reduction operator.
The overall running time is $\Oh{\alpha\log p}$.
For randomly permuted local inputs of size $n/p$, it is easy to show that our scheme yields a truthful estimator for the median, i.e., we get a result with expected rank $n/2$.
We conjecture that similar quality bounds as from  Dean et al.~\cite{dean2014} can be shown for our scheme, i.e., that we get rank $n/2(1\pm cn^{-\gamma})$ with high probability for some constant $c$ and $\gamma$.
Our experiments in Appendix~\ref{app:median selection trees experiments} indicate that the approximation guarantee of the binary tree is about $1.44n^{-0.39}$ whereas the approximation guarantee of the ternary tree is about $2n^{-0.37}$.

\section{Sorting Algorithms from Tiny to Huge Inputs}
\label{s:related}

\iftoggle{extended}{
  \begin{table}[b]
    \centering
    \caption{\label{t:time}Complexity of various parallel sorting algorithms. Implicit $\Oh{\cdot}$.}
    \tabcolsep=0.11cm
    \begin{tabular}{l|rrrr}
      Algorithm                                                  & Latency $[\alpha]$ & Comm. Vol. $[\beta]$      & Remarks                                  & Acronym      \\\hline\hline
      (All-)gather-merge                                         & $\log p$           & $n$                       & see Sect.~\ref{s:hypercube architecture} & \aagather    \\
      Fast Work-Inefficient Sort (\afir, \afis) \cite{Axtmann15} & $\log p$           & $n/\sqrt{p}$              & here: robust                             & \afis        \\\hline
      Bitonic sort~\cite{johnsson84}                             & $\log^2 p$         & $\frac{n}{p} \log^2 p$    & deterministic                            & \abitonic    \\\hline
      \aminisort \cite{SieWol11}                                 & $\log^2 p$         & $\log^2 p$                & $n=p$                                    &              \\
      Hypercube quicksort \cite{wagar87}                         & $\log^2 p$         & $\frac{n}{p} \log p$      & best case                                &              \\
      $+$median of medians \cite{Lan92}                          & $\log^2 p$         & $p+\frac{n}{p} \log p$    & average case                             &              \\
      Robust hypercube quicksort Sect.~\ref{s:quicksort}         & $\log^2 p$         & $\frac{n}{p} \log p$      & new, expected case                       & \aquick      \\\hline
      HykSort \cite{Sundar13}                                    & $\geq k\log_kp$    & $\geq \frac{n}{p}\log_kp$ & not robust                               & \ahyk        \\
      Adaptive Multilevel Samplesort (AMS) \cite{Axtmann15}      & $k\log_kp$         & $\frac{n}{p}\log_kp$      & here: robust                             & \aams        \\\hline
      Sample sort \cite{BleEtAl91}                               & $\geq p$           & $\geq n/p$                & $+$sampling cost                         & \asamplesort \\
      Multiway mergesort \cite{VarEtAl91,Axtmann15}              & $\geq p$           & $\geq n/p$                & $+$splitting cost
    \end{tabular}
  \end{table}
}{
  \begin{table}[b]
    \caption{\label{t:time}Complexity of various parallel sorting algorithms. Implicit $\Oh{\cdot}$.}
    \tabcolsep=0.11cm
    \begin{tabular}{l|rrr}
      Algorithm                                   & Latency $[\alpha]$ & Comm. Vol. $[\beta]$      & Remarks                        \\\hline\hline
      (All-)gather-merge                          & $\log p$           & $n$                       & \ref{s:hypercube architecture} \\
      \afis \cite{Axtmann15}                        & $\log p$           & $n/\sqrt{p}$              & here:robust                    \\\hline
      Bitonic \cite{johnsson84}                   & $\log^2 p$         & $\frac{n}{p} \log^2 p$    & deterministic                  \\\hline
      \aminisort \cite{SieWol11}                  & $\log^2 p$         & $\log^2 p$                & $n=p$                          \\
      HC quicksort \cite{wagar87}                 & $\log^2 p$         & $\frac{n}{p} \log p$      & best case                      \\
      $+$median of medians \cite{Lan92}           & $\log^2 p$         & $p+\frac{n}{p} \log p$    & average case                     \\
      HC quicksort Sect.~\ref{s:quicksort}        & $\log^2 p$         & $\frac{n}{p} \log p$      & here:robust                     \\\hline
      HykSort \cite{Sundar13}                     & $\geq k\log_kp$    & $\geq \frac{n}{p}\log_kp$ & not robust                     \\
      AMS \cite{Axtmann15}                        & $k\log_kp$         & $\frac{n}{p}\log_kp$      & here:robust                    \\\hline
      Sample sort \cite{BleEtAl91}                & $\geq p$           & $\geq n/p$                & $+$sampling cost               \\
      Multiway merges. \cite{VarEtAl91,Axtmann15} & $\geq p$           & $\geq n/p$                & 
    \end{tabular}
  \end{table}
}

We now outline a spectrum of parallel sorting algorithms (old and new), analyze them and use the result to compare their performance for different values of $n$ and $p$. Table~\ref{t:time} summarizes the results, roughly going from algorithms for small inputs to ones for larger and larger inputs. We only give the cost terms with respect to latency ($\alpha$) and communication volume ($\beta$) per PE because the relevant local work is always the same $\Oh{\frac{n}{p}\log n}$.
We now compare the algorithms with regard to running time and robustness. Appendices~\ref{app:related work} and~\ref{app:more related work} describe the algorithms in more detail.

We recently proposed Fast Work-Inefficient Ranking (\afir)~\cite{Axtmann15}, a simple but fast ranking algorithm for small inputs where the PEs are arranged in an array of size $\Oh{\sqrt{p}} \times \Oh{\sqrt{p}}$.
The result of the algorithm is that each PE knows the global rank of all input elements in its row. In theory this is a very good algorithm for $n=\Oh{\sqrt{p}}$ since we have only logarithmic delay.
In Section~\ref{s:fastInefficient ams} we propose Robust fast work-inefficient sort (\afis) which makes the algorithm robust against identical keys and converts its output to a classical sorted permutation of the input.


For small inputs the running
time of Bitonic sort~\cite{Batcher68,johnsson84} is dominated by $\log^2p$ startup overheads.  This gives it a similar running time as the parallel quicksort algorithms to be
discussed next.  However, for $n=\omega(p\alpha/\beta)$ the term
$\beta \frac{n}{p} \log^2 p$ dominates -- all the data is exchanged $\log^2
p$ times.
This makes it unattractive for large inputs compared to
quicksort and other algorithms designed for large inputs.

There are many parallel variants of quicksort. Hypercube quicksort uses the hypercube communication pattern, is simple and can exploit locality of the network.
In the simplest original form by Wagar \cite{wagar87}, PE 0 uses its local median as a pivot.
This is not robust against skew and even for average case inputs it only works for rather large $n$.
One can make this more robust for average case inputs by using the global median of the local medians \cite{Lan92}. 
However, this introduces an additional $\beta p$ term and leads to non-polylogarithmic execution time.
Therefore in Section~\ref{s:quicksort} we make quicksort robust against skew and propose to use a median selection algorithm that is both fast and accurate.
Non-hypercube distributed memory quicksort leads to more irregular communication patterns and has its own load balancing problems because we cannot divide PEs fractionally.
Siebert and Wolf \cite{SieWol11} exploit the special case $n=p$ where this is no problem.


Quicksort exchanges the data at least $\log p$ times.
More general approaches partition the input into $k$ parts on each recursion level.
Those algorithms move the data $\Oh{\log_kp}$ times and the lower bound of the running time is
$$\OmL{\frac{n}{p}\log n}+\beta\frac{n}{p}\log_kp+\alpha k\log_kp\punkt$$
Many algorithms have been devised for the special case $k=p$ ~\cite{BleEtAl91, SolKal10}.
These single level algorithms exchange data only once.
Sample sort~\cite{BleEtAl91} achieves the desired bound for relatively large inputs of $n=\Om{p^2/\log p}$.
Helman et al. \cite{Helman98} propose initial random shuffling to make samplesort robust against skewed input distributions (this was already proposed in \cite{SanHan97e}).
One can even achieve perfect partitioning by finding optimal splitters in a multiway mergesort approach \cite{VarEtAl91,SSP07,Axtmann15}.
This requires $n=\Om{p^2\log p}$ for achieving the desired bound.
Gerbessiotis and Valiant \cite{GerVal94} develop a multilevel sample sort for the BSP model \cite{Val94}.
However, implementing the data exchange primitive of BSP requires $p$ startup overheads in each round.
Cole and Ramachandran \cite{ColRam10} and Blelloch et al. \cite{BGSV10} give multi-level sorting algorithms for cache oblivious shared memory models.
Translating them to a distributed memory model appears nontrivial and is likely to result in worse bounds.

Sundar et al.~\cite{Sundar13} develop and implement \emph{HykSort}, a multi-level generalization of hypercube quicksort. HykSort is ``almost'' robust against skew. However, in the worst case, data may become severely imbalanced. The paper mentions measures for solving this problem, but does not analyze them. Furthermore, HykSort uses the operations {\tt MPI\_Comm\_Split} whose current implementations need time $\Om{\beta p}$. This is why we put a ``$\geq$'' in the corresponding row of Table~\ref{t:time}. HykSort is also not robust with respect to duplicate keys.

Our recently published $k$-way sample sort algorithm, \emph{AMS-sort}~\cite{Axtmann15}, guarantees a maximum imbalance of $\epsilon \frac{n}{p}$ for arbitrary inputs.
AMS-sort avoids the data imbalances within target subcubes that may make HykSort inefficient. Our paper~\cite{Axtmann15} describes (but does not implement) a way to select communication partners such that each PE sends and receives $\Theta(k)$ messages for a $k$-way data exchange. 
Similarly, a tie-breaking technique is proposed (but not implemented) which simulates unique keys with very low overhead.
In Appendix~\ref{app:ams} we describe \aams, our robust implementation of AMS-sort (see also Section~\ref{s:fastInefficient ams}).
Note that the running time of the multi-level algorithms described above with $\Th{\log p}$ levels is $\Om{\log^2 p}$.
For example, the running time of AMS-sort is then $\Oh{\log^2 p \log \log p}$.
However, in this case quicksort is the simpler algorithm, has a startup latency of $\Oh{\log^2 p}$ and smaller constant startup latency factors.

\section{Robust Fast Work-Inefficient Sorting and Multi-Level Sample Sort}\label{s:fastInefficient ams}

We propose a fast sorting algorithm (\afis) for sparse and very small inputs
that is robust against duplicate keys.
Our algorithm computes the rank of each element in the global data.
Then, optionally, a data delivery routine sends the elements to the appropriate PEs according to their ranks.
For ranking the input data, we use the algorithm we recently proposed~\cite{Axtmann15} (see also~\cite{IKS09} and Appendix~\ref{app:fast work ranking}).
Unfortunately the proposed approach calculates the same rank for equal elements.
However, our data delivery technique requires unique ranks in the range of $0..n-1$ as input to achieve balanced output.
Consider a tie-breaking policy that logically treats
an element as a quadruple $(x,r,c,i)$ where $x$ is the key, 
$r$ the row, $c$ the column, and $i$ the position in the locally sorted data.
We can then define a total ordering between these quadruples as their lexicographical order. 
We now exploit the fact that we do not rely on the true column $c$ (row $r$) but on the information weather an element came from the left (above) or from the right (below).
A modified all-gather-merge and a modified compare function implement that policy without any communication overhead for communicating the $(r,c,i)$ information.
Appendix~\ref{app:fastInefficient} describes this policy in more detail.
After computing the ranks of the input elements, we have to redistribute them to get sorted output. The element with rank $i$ is mapped to PE $ip/n$.
We now exploit the fact that each PE column stores the complete ranked input. We thus can afford to discard all elements that are not mapped to the same column without losing any element. The data delivery problem now can be solved locally in each column. We use a hypercube algorithm for routing the elements to their correct row. In the $i$-th iteration of this algorithm, at most
 $n/(p^{\frac{1}{2}} 2^i)$ elements have to be delivered in their respective subcube. Even if all these elements are concentrated on the same PE, this requires at most $\alpha+\beta n/(p^{\frac{1}{2}} 2^i)$ time.
Summing this over all iterations yields overall time
$\Oh{\alpha\log p + \beta \frac{n}{\sqrt{p}}}$, i.e., the asymptotical running time is the same for just ranking or ranking plus data delivery.

The prototypical implementation of AMS in \cite{Axtmann15} does not ensure that each PE communicates with at most $\Oh{k}$ PEs in each
level of recursion \frage{ps added:}(deterministic message assignment). Similarly, \cite{Axtmann15} proposes but does not implement 
a tie-breaking scheme to become robust against duplicate keys with almost no communication overhead.
in Appendix~\ref{app:ams}, we present a complete implementation of AMS-sort (\aams) including both improvements outlined above.

\section{Robust Quicksort on Hypercubes}\label{s:quicksort}
\begin{algorithm}[t]
  \begin{algorithmic}
    \caption{Robust Quicksort on Hypercubes}\label{alg:random quicksort}
    \State \textbf{Input:} $a = \{ a_1,\ldots,a_{n/p} \}$ a set of local input elements, $p = 2^{d}$ number of PEs, PE number $i$
    \State $a \gets$ randomly redistribute $a$\Comment{see Section~\ref{s:randomShuffle}}
    \State \Call{Sort}{$a$}
    \For{$j\Is d - 1$ {\bf downto} 0} \Comment{iterate over cube dims}
      \State $s \gets$ calculate splitter in parallel \Comment{see Section~\ref{s:approxRankSelection}}
      \If{\Call{isEmpty}{$s$}} \Return $a$ \Comment{no elements in cube}
      \EndIf 
      \State split $a$ into $L\cdot R$\Comment{see text}
      \State $i'\Is i\oplus 2^j$ \Comment{communication partner}
      \If{$i'<i$}
        \State send $L$ to $i'$ and receive $R'$ from $i'$
        \State $a\gets$ merge $R$ with $R'$
      \Else
        \State send $R$ to $i'$ and receive $L'$ from $i'$
        \State $a\gets$ merge $L$ with $L'$
      \EndIf
    \EndFor
    \State \Return $a$
  \end{algorithmic}
\end{algorithm}

Previous implementations of hypercube-based quicksort (e.g. \cite{wagar87, Lan92}) have three major drawbacks.
First, they are not robust against duplicates as no tie-breaking is performed.
Second, the algorithms do not consider non-uniform input distributions. 
It can be shown (e.g. \cite{Lei92})
that skewed inputs exist where the imbalance increases rapidly over the first $\frac{\log p}{2}$ recursions.
After $\frac{\log p}{2}$ recursions, $\ceiling{\sqrt{p}}$ PEs hold $\flooring{n/ \sqrt{p}}$ elements each. 
This bound even applies for unique input keys and true medians as a global splitter.
Third, a fast high quality splitter selection with median approximation guarantees is essential.
Otherwise, quicksort gets impractical for large $p$ as the load imbalance accumulates or the splitter selection dominates the running time for small inputs.
Previous quicksort implementations use one of two different approaches to calculate the splitter.
The first approach~\cite{wagar87} selects a random PE which just broadcasts its local median.
The second approach~\cite{Lan92} gathers the local median of each PE and calculates the median of medians.

We propose a simple but robust hypercube quicksort that overcomes these problems.
Algorithm~\ref{alg:random quicksort} gives pseudocode.
First, our algorithm randomly redistributes the input (see Section~\ref{s:randomShuffle}).
Then the data is sorted locally.
The main loop goes through the dimensions of the hypercube, highest dimension first. In iteration $j$, the fast median selection algorithm from Section~\ref{s:approxRankSelection} is used to approximate the median of the data in each $j+1$-dimensional subcube. This key is used as a splitter $s$ for that subcube. For robustness against repeated keys, we use a low overhead tie-breaking scheme that works locally without the need to communicate additional information: Suppose
a PE holds a sorted sequence $a$ of the form 
$a=a_{\ell}\cdot s^m\cdot a_r$, i.e., the splitter $s$ appears $m$ times locally. 
We can split $a$ into two subsequences $L=a_{\ell}\cdot s^x$ and $R=s^{m-x}\cdot a_r$ where we are free to choose $x\in 0..m$. We choose $x$ in such a way that $|a_{\ell}s^x|$ is as close to $|a|/2$ as possible. In a communication step, PEs differing in bit $j$ of their PE number exchange these sequences so that the PE with the 0-bit gets the
two $L$ sequences and the PE with the 1-bit gets the $R$ sequences. These two sequences are then merged. 

\subsubsection*{Analysis}

Here we give evidence why we believe the following conjecture to be true: 
\begin{conjecture}
Our robust hypercube quicksort runs in expected time 
$$\OhL{\frac{n}{p}\log n+\beta\frac{n}{p}\log p +\alpha\log^2 p}$$
for arbitrary inputs with unique keys.
\end{conjecture}
The proof is straight-forward for best-case inputs where all sequences are split perfectly in each step -- we have $\Oh{\frac{n}{p}\log\frac{n}{p}}$ time for local sorting and time $\Oh{\beta\frac{n}{p}\log p +\alpha\log p}$ for random permutation (note that this is small compared to the overall execution time for $n\ll \frac{\alpha}{\beta}\log p$). Each iteration of the main loop costs time $\Oh{\alpha\log p}$ for splitter selection and time $\Oh{\beta\frac{n}{p}}$ for local data exchange and merging.

The initial random shuffling is pivotal for generalizing the result to worst case inputs. From a general point of view this is not surprising since it essentially transforms a worst case input into an average case input. A closer look shows that it simplifies the algorithm in three different respects:

The first advantage is that this makes the algorithm robust against skew also in intermediate steps. 
To see this, consider the path an input element travels during the execution of the algorithm. It travels from a random position (determined by the random shuffle) to its final destination (governed by the splitters used in the $\log p$ levels of recursion). \anmerkung{Element placed at random PEs within the subcube.}Hence, the average case analysis of greedy hypercube packet routing (e.g., \cite{Lei92}) applies that shows that overall contention is low with high probability.

The second advantage is that random shuffle also makes it possible to quickly find a good approximation of the median using the algorithm from Section~\ref{s:approxRankSelection}. To see that this also implies good overall load balance, consider the following calculation.
According to Section~\ref{s:approxRankSelection}, the
imbalance factor introduced in recursion level $i$ is bounded by 
$1+(n/2^i)^{-\gamma}$ for some positive constant $\gamma$.
We use $n/2^i$ rather than the actual (already imbalanced) subproblem size here because, at the end, the PEs with highest load determine the overall imbalance. Hence, the most highly loaded subproblems are essential here. For those, $n/2^i$ is an underestimate of the subproblem size which results in an \emph{overestimate} of the imbalance factor introduced by splitting that subproblem. Now we can estimate the overall imbalance factor $I$ as
\begin{align*}
I & = \prod_{i<\log p}    
         \left(1+\left(\frac{2^i}{n}\right)^{\gamma}\right)
    = e^{\ln\prod_{i<\log p}
         \left(1+\left(\frac{2^i}{n}\right)^{\gamma}\right)}\\
  & = e^{\sum_{i<\log p}           
       \ln\left(1+\left(\frac{2^i}{n}\right)^{\gamma}\right)}
  \leq e^{\sum_{i<\log p}           
       \left(\frac{2^i}{n}\right)^{\gamma}}\\
  & = e^{ n^{-\gamma}\sum_{i<\log p}                  
         (2^{\gamma})^{i}}
  \leq e^{\frac{\left(\frac{p}{n}\right)^{\gamma}}{2^{\gamma}-1}}
    = \Oh{1} \text{ for $n\geq p$}\punkt
\end{align*}
The first ``$\leq$'' uses the Taylor series development of $\ln$ and the second one is a geometric sum.
This calculation also shows that the overall number of elements in any subcube in any level of recursion is balanced.
We argue that the initial random data distribution also implies a random data distribution of the elements in each subcube. Therefore, the overall balance of the number of elements also implies that the individual number of elements in each PE is not too imbalanced.%
\footnote{A full proof may not be trivial however. For example, there \emph{are} significant fluctuations in the individual PE loads for $n/p=\Oh{\log p}$ . However, these are likely to only introduce overhead factors dwarfed by the overhead for median selection as elements are randomly placed on the PEs associated with the subcube. We also have to take into accout that the probability of a maximal imbalance factor of $1+(n/2^i)^{-\gamma}$, introduced in recursion level $i$, decreases as the load of subcubes decreases.\todo{We might give a benchmark which tells us that this is not a problem we have to think about!}}

Finally, random data shuffling also helps our simple local tie-breaking -- each PE has a random sample of the globally available data and hence our local balancing approximates a global balancing.
  
\section{Experimental Results}\label{s:experiments}

\def\RunnintimeLegend{%
  \legend{
    \aams,
    \abitonic,
    \afis,
    \agather,
    \aallgather,
    \aquick,
    \ahyk
  }%
}

\pgfplotscreateplotcyclelist{my exotic}{%
teal,every mark/.append style={fill=teal!80!black},mark=star\\%
orange,every mark/.append style={fill=orange!80!black},mark=triangle*\\%
brown!60!black,every mark/.append style={fill=brown!80!black},mark=square*\\%
red!70!white,mark=star\\%
lime!80!black,every mark/.append style={fill=lime},mark=diamond*\\%
red,densely dashed,every mark/.append style={solid,fill=red!80!black},mark=*\\%
yellow!60!black,densely dashed,every mark/.append style={solid,fill=yellow!80!black},mark=square*\\%
black,every mark/.append style={solid,fill=gray},mark=otimes*\\%
blue,densely dashed,mark=star,every mark/.append style=solid\\%
red,densely dashed,every mark/.append style={solid,fill=red!80!black},mark=diamond*\\%
}
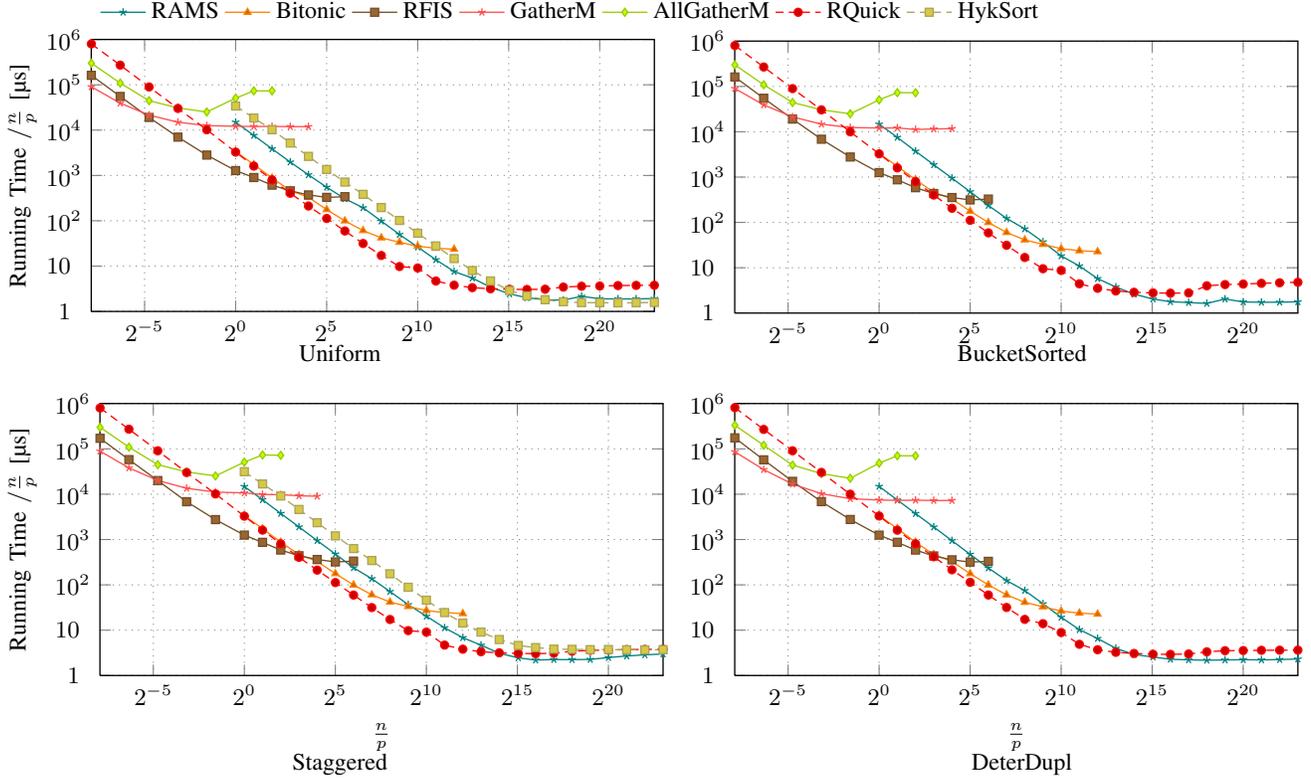
\begin{figure*}[t]\centering
\captionsetup[subfloat]{labelformat=empty, position=top}
\pgfplotsset{
  plotconfig/.style={
    mark size=1.5pt,
    cycle list name=my exotic,
    height=52mm,
    width=\textwidth,
    xtickten={-5, 0, 5, 10, 15, 20},
    legend style={
      fill=none,
      draw=none}
  }
}

\captionsetup[subfigure]{labelformat=empty}
\begin{subfigure}[b]{0.5\textwidth}
\begin{tikzpicture}
  \begin{axis}[
    xmax=8388608,
    xmin=0.0041,
    ymin=1,
    ymax=1000000,
    ylabel={Running Time $/\frac{n}{p}$ [\textmu s]},
    plotconfig,
    xlabel near ticks,
    legend style={at={(0,1.1)}, anchor=west},
    legend columns=7, 
    log base x=2,
    log base y=10,
    xmode=log,
    ymode=log,
    line width=0.5pt,
    tick style={
      line cap=round,
      thin,
      major tick length=4pt,
      minor tick length=2pt,
    },
  major grid style={thin,dotted,color=black!50},
  minor grid style={thin,dotted,color=black!50},
  grid,
  font=\small,
  compat=1.3,
  ytickten={0, 1, 2, 3, 4, 5, 6},
  yticklabels={$1\:\:$, $10\:\:$, $10^2$, $10^3$, $10^4$, $10^5$, $10^6$},
  y label style={at={(-0.08,0.5)}},],
    ]
    \addplot coordinates { (1,14988.2) (2,7637.1) (4,3869.8) (8,1976.975) (16,1034.1) (32,550.225) (64,306.759) (128,193.886) (256,98.4172) (512,49.8891) (1024,26.2828) (2048,13.8214) (4096,7.60249) (8192,5.44888) (16384,3.42112) (33768,2.44039) (65536,2.00316) (131072,1.81123) (262144,1.77874) (524288,2.15411) (1048576,1.90405) (2097152,1.88671) (4194304,1.89073) (8388608,1.92644) };
    \addlegendentry{algo=AMS};
    \addplot coordinates { (1,3386.4) (2,1739.6) (4,878) (8,465.05) (16,329.475) (32,176.6625) (64,99.7281) (128,60.8891) (256,42.0086) (512,33.5941) (1024,27.4631) (2048,24.8105) (4096,23.5895) };
    \addlegendentry{algo=Bitonic};
    \addplot coordinates { (0.00411523,161886.6) (0.0123457,55323) (0.037037,19040.4) (0.111111,7003.8) (0.333333,2815.8) (1,1280.5) (2,899.8) (4,611.5) (8,457.825) (16,372.175) (32,328.294) (64,339.909) };
    \addlegendentry{algo=FIS};
    \addplot coordinates { (0.00411523,90444.6) (0.0123457,39382.2) (0.037037,21470.4) (0.111111,14945.4) (0.333333,12574.2) (1,12150) (2,12031.4) (4,12053) (8,11943.5) (16,11927.45) };
    \addlegendentry{algo=Gather};
    \addplot coordinates { (0.00411523,299570.4) (0.0123457,109139.4) (0.037037,44604) (0.111111,30866.4) (0.333333,25163.4) (1,50742) (2,73426.7) (4,73514.2) };
    \addlegendentry{algo=GatherAll};
    \addplot coordinates { (0.00411523,799032.6) (0.0123457,271431) (0.037037,89656.2) (0.111111,30092.4) (0.333333,10179) (1,3273.5) (2,1607.7) (4,800.35) (8,405.05) (16,211.85) (32,112.5) (64,59.5594) (128,31.5375) (256,17.1617) (512,9.78828) (1024,9.08809) (2048,4.68193) (4096,3.79829) (8192,3.34761) (16384,3.15979) (33768,3.08318) (65536,3.05049) (131072,3.09168) (262144,3.44232) (524288,3.59712) (1048576,3.6329) (2097152,3.72114) (4194304,3.7564) (8388608,3.78569) };
    \addlegendentry{algo=Quick};
    \addplot coordinates { (1,34035.2) (2,18522.5) (4,10192.25) (8,5183.475) (16,2635.325) (32,1353.86) (64,713.809) (128,383.131) (256,196.145) (512,101.679) (1024,53.052) (2048,27.8553) (4096,14.6586) (8192,7.99097) (16384,4.70055) (33768,2.90126) (65536,2.20424) (131072,1.81427) (262144,1.63562) (524288,1.56533) (1048576,1.55878) (2097152,1.54424) (4194304,1.55195) (8388608,1.57834) };
    \addlegendentry{algo=ZHyk};
    \RunnintimeLegend
    
    \RunnintimeLegend
\end{axis}
\end{tikzpicture}
\vspace{-0.7cm}
\caption{Uniform}
\end{subfigure}%
\begin{subfigure}[b]{0.5\textwidth}
\begin{tikzpicture}
  \begin{axis}[
    xmax=8388608,
    xmin=0.0041,
    ymin=1,
    ymax=1000000,
    plotconfig,
    ylabel near ticks,
    xlabel near ticks,
    legend columns=1, 
    log base x=2,
    log base y=10,
    xmode=log,
    ymode=log,
    line width=0.5pt,
    tick style={
      line cap=round,
      thin,
      major tick length=4pt,
      minor tick length=2pt,
    },
  major grid style={thin,dotted,color=black!50},
  minor grid style={thin,dotted,color=black!50},
  grid,
  font=\small,
  ytickten={0, 1, 2, 3, 4, 5, 6},
  yticklabels={$1\:\:$, $10\:\:$, $10^2$, $10^3$, $10^4$, $10^5$, $10^6$},
    ]
    \addplot coordinates { (1,14737.6) (2,7523.9) (4,3732.15) (8,1871.275) (16,948.1375) (32,474.994) (64,236.541) (128,122.811) (256,71.9547) (512,37.1285) (1024,18.2229) (2048,10.9219) (4096,5.76406) (8192,3.76201) (16384,2.61993) (33768,2.04162) (65536,1.78022) (131072,1.70583) (262144,1.63176) (524288,2.04201) (1048576,1.75928) (2097152,1.72983) (4194304,1.74023) (8388608,1.76928) };
    \addlegendentry{algo=AMS};
    \addplot coordinates { (1,3327.8) (2,1736.4) (4,875.35) (8,463.175) (16,330.3625) (32,176.75) (64,99.7406) (128,60.0953) (256,41.2023) (512,32.8242) (1024,26.3537) (2048,23.7745) (4096,22.6314) };
    \addlegendentry{algo=Bitonic};
    \addplot coordinates { (0.00411523,161400.6) (0.0123457,54820.8) (0.037037,19051.2) (0.111111,6843.6) (0.333333,2775) (1,1255.4) (2,871.5) (4,583.7) (8,434.35) (16,353.9) (32,313.744) (64,327.022) };
    \addlegendentry{algo=FIS};
    \addplot coordinates { (0.00411523,90007.2) (0.0123457,39204) (0.037037,21254.4) (0.111111,14722.2) (0.333333,12372.6) (1,12101.6) (2,12136.8) (4,11231.85) (8,11524.2) (16,11721.6) };
    \addlegendentry{algo=Gather};
    \addplot coordinates { (0.00411523,298987.2) (0.0123457,108847.8) (0.037037,44296.2) (0.111111,30540.6) (0.333333,24906.6) (1,50566.6) (2,73002.6) (4,72634.6) };
    \addlegendentry{algo=GatherAll};
    \addplot coordinates { (0.00411523,799761.6) (0.0123457,268223.4) (0.037037,89591.4) (0.111111,30362.4) (0.333333,9966) (1,3255.4) (2,1594.3) (4,797.2) (8,401.75) (16,205.4) (32,111.4375) (64,58.84375) (128,31.3016) (256,16.8414) (512,9.52031) (1024,8.74238) (2048,4.42393) (4096,3.52197) (8192,3.07578) (16384,2.87469) (33768,2.78349) (65536,2.74769) (131072,2.78366) (262144,3.99865) (524288,4.2538) (1048576,4.3643) (2097152,4.53038) (4194304,4.65936) (8388608,4.78229) };
    \addlegendentry{algo=Quick};
    \legend{}
  \end{axis}
\end{tikzpicture}
\vspace{-0.3cm}
\caption{BucketSorted}
\end{subfigure}
\vspace{-0.4cm}
\begin{subfigure}[b]{0.5\textwidth}
\begin{tikzpicture}
  \begin{axis}[
    xmax=8388608,
    xmin=0.0041,
    ymin=1,
    ymax=1000000,
    plotconfig,
    ylabel near ticks,
    xlabel near ticks,
    ylabel={Running Time $/\frac{n}{p}$ [\textmu s]},
    legend columns=1, 
    log base x=2,
    log base y=10,
    xmode=log,
    ymode=log,
    xlabel={$\frac{n}{p}$},
    line width=0.5pt,
    tick style={
      line cap=round,
      thin,
      major tick length=4pt,
      minor tick length=2pt,
    },
  major grid style={thin,dotted,color=black!50},
  minor grid style={thin,dotted,color=black!50},
  grid,
  font=\small,
  ytickten={0, 1, 2, 3, 4, 5, 6},
  yticklabels={$1\:\:$, $10\:\:$, $10^2$, $10^3$, $10^4$, $10^5$, $10^6$},
    ]
    \addplot coordinates { (1,14772.4) (2,7542.4) (4,3758.25) (8,1896.525) (16,947.4125) (32,478.031) (64,238.634) (128,136.692) (256,71.2242) (512,36.6273) (1024,20.3287) (2048,11.2799) (4096,6.84116) (8192,4.67314) (16384,3.10605) (33768,2.45269) (65536,2.19195) (131072,2.25198) (262144,2.24042) (524288,2.26489) (1048576,2.48671) (2097152,2.69476) (4194304,2.85932) (8388608,2.98552) };
    \addlegendentry{algo=AMS};
    \addplot coordinates { (1,3362.2) (2,1736.3) (4,873.4) (8,467.025) (16,329.5375) (32,177.081) (64,99.275) (128,60.5125) (256,41.7078) (512,33.3359) (1024,27.1322) (2048,24.4443) (4096,23.22) };
    \addlegendentry{algo=Bitonic};
    \addplot coordinates { (0.00411523,171217.8) (0.0123457,57915) (0.037037,19990.8) (0.111111,6829.2) (0.333333,2750.4) (1,1250.8) (2,866.3) (4,586.2) (8,438.35) (16,362.8125) (32,320.256) (64,334.316) };
    \addlegendentry{algo=FIS};
    \addplot coordinates { (0.00411523,89181) (0.0123457,38021.4) (0.037037,20077.2) (0.111111,13491) (0.333333,11128.8) (1,10678.3) (2,9839.6) (4,9758.75) (8,9229.875) (16,9017.475) };
    \addlegendentry{algo=Gather};
    \addplot coordinates { (0.00411523,299278.8) (0.0123457,109155.6) (0.037037,44749.8) (0.111111,31026.6) (0.333333,25293) (1,50995.2) (2,73408) (4,72068.1) };
    \addlegendentry{algo=GatherAll};
    \addplot coordinates { (0.00411523,798643.8) (0.0123457,271447.2) (0.037037,90703.8) (0.111111,30047.4) (0.333333,10129.2) (1,3278.8) (2,1609.6) (4,802.6) (8,405.175) (16,211.4625) (32,112.331) (64,59.5344) (128,31.5672) (256,17.2117) (512,9.78203) (1024,9.05059) (2048,4.68965) (4096,3.80562) (8192,3.34712) (16384,3.16926) (33768,3.07513) (65536,3.0284) (131072,3.10518) (262144,3.43291) (524288,3.59225) (1048576,3.63986) (2097152,3.72729) (4194304,3.76777) (8388608,3.79867) };
    \addlegendentry{algo=Quick};
    \addplot coordinates { (1,31335) (2,16844.4) (4,9146.6) (8,4601.65) (16,2350.99) (32,1203.38) (64,632.156) (128,344.291) (256,175.519) (512,88.534) (1024,45.925) (2048,24.446) (4096,14.3356) (8192,9.06995) (16384,6.21598) (33768,4.60102) (65536,4.12946) (131072,3.81614) (262144,3.77993) (524288,3.72084) (1048576,3.74269) (2097152,3.68478) (4194304,3.67278) (8388608,3.70924) };
    \addlegendentry{algo=ZHyk};
    
    \legend{}
  \end{axis}
\end{tikzpicture}
\vspace{-0.3cm}
\caption{Staggered}
\end{subfigure}%
\begin{subfigure}[b]{0.5\textwidth}
\begin{tikzpicture}
  \begin{axis}[
    xmax=8388608,
    xmin=0.0041,
    ymin=1,
    ymax=1000000,
    plotconfig,
    xlabel near ticks,
    legend columns=1, 
    log base x=2,
    log base y=10,
    xmode=log,
    ymode=log,
    xlabel={$\frac{n}{p}$},
    line width=0.5pt,
    tick style={
      line cap=round,
      thin,
      major tick length=4pt,
      minor tick length=2pt,
    },
  major grid style={thin,dotted,color=black!50},
  minor grid style={thin,dotted,color=black!50},
  grid,
  font=\small,
  compat=1.3,
  ytickten={0, 1, 2, 3, 4, 5, 6},
  yticklabels={$1\:\:$, $10\:\:$, $10^2$, $10^3$, $10^4$, $10^5$, $10^6$},
  y label style={at={(-0.08,0.5)}},]
    ]
    \addplot coordinates { (1,14865.2) (2,7467.3) (4,3739.4) (8,1888.875) (16,945.075) (32,471.4875) (64,235.669) (128,124.269) (256,74.9516) (512,38.2809) (1024,19.2436) (2048,10.2771) (4096,6.56118) (8192,3.98289) (16384,2.90751) (33768,2.55699) (65536,2.28417) (131072,2.21402) (262144,2.17259) (524288,2.20433) (1048576,2.23394) (2097152,2.21017) (4194304,2.24856) (8388608,2.30946) };
    \addlegendentry{algo=AMS};
    \addplot coordinates { (1,3399.2) (2,1741) (4,875.4) (8,467.175) (16,328.3125) (32,176.456) (64,99.6375) (128,60.1641) (256,41.2258) (512,32.834) (1024,26.3658) (2048,23.7931) (4096,22.6478) };
    \addlegendentry{algo=Bitonic};
    \addplot coordinates { (0.00411523,174765.6) (0.0123457,57250.8) (0.037037,19197) (0.111111,6859.8) (0.333333,2764.8) (1,1249.6) (2,868.8) (4,585.7) (8,433.925) (16,356.6875) (32,317.556) (64,329.994) };
    \addlegendentry{algo=FIS};
    \addplot coordinates { (0.00411523,86216.4) (0.0123457,34878.6) (0.037037,16723.8) (0.111111,10251) (0.333333,7951.2) (1,7434.7) (2,7311.2) (4,7364.05) (8,7259.375) (16,7263.94) };
    \addlegendentry{algo=Gather};
    \addplot coordinates { (0.00411523,332132.4) (0.0123457,120009.6) (0.037037,43945.2) (0.111111,28440) (0.333333,22431) (1,48377.3) (2,70316.3) (4,70414.8) };
    \addlegendentry{algo=GatherAll};
    \addplot coordinates { (0.00411523,812446.2) (0.0123457,268741.8) (0.037037,90752.4) (0.111111,30261.6) (0.333333,9999) (1,3301.9) (2,1605.9) (4,807.15) (8,416.675) (16,214.6375) (32,113.8) (64,59.6875) (128,31.7375) (256,17.2984) (512,13.9176) (1024,8.878125) (2048,4.87998) (4096,3.70444) (8192,3.24812) (16384,3.06749) (33768,2.9456) (65536,2.93082) (131072,2.98731) (262144,3.33098) (524288,3.52665) (1048576,3.55115) (2097152,3.59611) (4194304,3.61697) (8388608,3.63169) };
    \addlegendentry{algo=Quick};
    \legend{}
  \end{axis}
\end{tikzpicture}
\vspace{-0.3cm}
\caption{DeterDupl}
\end{subfigure}
\vspace{-0.3cm}
\caption{Running times of each algorithm on 262\,144 cores. \ahyk crashes on input instances \deterdupl and \bucketsorted.}
\label{fig:largest instance rt}
\end{figure*}

\pgfplotscreateplotcyclelist{my exotic}{%
teal,every mark/.append style={fill=teal!80!black},mark=star\\%
orange,every mark/.append style={fill=orange!80!black},mark=triangle*\\%
brown!60!black,every mark/.append style={fill=brown!80!black},mark=square*\\%
red!70!white,mark=star\\%
lime!80!black,every mark/.append style={fill=lime},mark=diamond*\\%
red,densely dashed,every mark/.append style={solid,fill=red!80!black},mark=*\\%
yellow!60!black,densely dashed,every mark/.append style={solid,fill=yellow!80!black},mark=square*\\%
black,every mark/.append style={solid,fill=gray},mark=otimes*\\%
blue,densely dashed,mark=star,every mark/.append style=solid\\%
red,densely dashed,every mark/.append style={solid,fill=red!80!black},mark=diamond*\\%
}
\begin{figure*}[t]\centering
\captionsetup[subfloat]{
  position=top}
\pgfplotsset{
  grid=both,
  major grid style={thin,dotted,color=black!50},
  minor grid style={thin,dotted,color=black!50},
  plotconfig/.style={
    mark size=1.5pt,
    cycle list name=my exotic,
    height=54mm,
    width=\textwidth
}
}

\begin{subfigure}[b]{0.5\textwidth}
\caption{\aquick vs. \antbquick}
\vspace{-0.2cm}
\begin{tikzpicture}
  \begin{axis}[
    xmin=1,
    xmax=1048576,
    ylabel style={align=center},
    ylabel={Running Time Ratio\\$[t_{\text{\aquick}} / t_{\text{\antbquick}}]$},
    legend style={
      at={(0.5,0.97)},anchor=north},
    legend columns=3,
    plotconfig,
    ymax=2.4,
    ymin=-0.1,
    xmode=log,
    log base x=2,
    y label style={at={(0.04,0.5)}},
    ytick={0, 0.5, 1, 1.5, 2},
    xtickten={0, 5, 10, 15, 20},
  ]

    \addplot coordinates { (1,1.10932) (2,1.08304) (4,1.08136) (8,1.10042) (16,1.08507) (32,1.11839) (64,1.12345) (128,1.1483) (256,1.16454) (512,1.2029) (1024,1.19517) (2048,1.38309) (4096,1.46865) (8192,1.54923) (16384,1.59654) (33768,1.64281) (65536,1.63908) (131072,1.64512) (262144,1.6162) (524288,1.6165) (1048576,1.63351) };
    \addlegendentry{\uniform};  
    \addplot coordinates { (1,0.461171) (2,0.311781) (4,0.233049) (8,0.184477) (16,0.145841) (32,0.112249) (64,0.0804964) };
    \addlegendentry{\bucketsorted};
    \addplot coordinates { (1,1.12793) (2,1.09097) (4,1.07729) (8,1.05181) (16,1.04552) (32,1.06574) (64,1.06175) (128,1.06797) (256,0.80002) (512,0.70801) (1024,0.941403) (2048,0.74562) (4096,0.765948) (8192,0.729456) (16384,0.541256) (33768,0.665349) (65536,0.56937) (131072,0.508018) (262144,0.545542) (524288,0.541192) (1048576,0.528647) };
    \addlegendentry{\staggered};
    \addplot coordinates { (1,0.607273) (2,0.421213) (4,0.308864) (8,0.236239) (16,0.187807) (32,0.148443) (64,0.116086) (128,0.0790322) };
    \addlegendentry{\deterdupl};
    \addplot coordinates { (1,1.09078) (2,1.0223) (4,0.99555) (8,0.984138) (16,0.967257) (32,0.913819) (64,0.791513) (128,0.65761) (256,0.50828) (512,0.367662) (1024,0.348179) (2048,0.242128) (4096,0.204714) (8192,0.169399) (16384,0.136003) (33768,0.111348) };
    \addlegendentry{\mirrored};

    \addplot[color=black] coordinates { (1,1) (1048576,1) };
\end{axis}
\end{tikzpicture}
\label{fig:quick robust}
\end{subfigure}%
\begin{subfigure}[b]{0.5\textwidth}

\caption{\aams vs. \antbams}
\vspace{-0.2cm}
\begin{tikzpicture}
  \begin{axis}[
    xmin=1,
    xmax=1048576,
    ylabel={[$t_{\text{\aams}} / t_{\text{\antbams}}$]},
    legend style={
      cells={align=left},
      at={(0.5,0.97)},anchor=north},
    legend columns=2,
    plotconfig,
    ymax=2.4,
    ymin=-0.1,
    xmode=log,
    log base x=2,
    ytick={0, 0.5, 1, 1.5, 2},
    y label style={at={(0.04,0.5)}},
    xtickten={0, 5, 10, 15, 20},
    ]
    \addplot coordinates { (1,1.15517) (2,1.1455) (4,1.15091) (8,1.14438) (16,1.13899) (32,1.07536) (64,1.0719) (128,1.07294) (256,1.12206) (512,1.10933) (1024,1.112) (2048,1.11687) (4096,1.09439) (8192,1.06987) (16384,1.03594) (33768,1.01987) (65536,1.01107) (131072,1.01078) (262144,0.998636) (524288,1.0039) (1048576,1.00184) };
    \addlegendentry{\uniform};
    \addplot coordinates { (1,0.892529) (2,0.937911) (4,0.846839) (8,0.69861) (16,0.448369) (32,0.081396) (64,0.0387227) (128,0.0237574) (256,0.0114082) (512,0.00605509) (1024,0.00335287) (2048,0.0017145) (4096,0.00124269) (8192,1.04033) (16384,1.01988) };
    \addlegendentry{\bucketsorted};
    \addplot coordinates { (1,1.14334) (2,1.13925) (4,1.14234) (8,1.14315) (16,1.13701) (32,1.14082) (64,1.13666) (128,1.13434) (256,1.12598) (512,1.11511) (1024,1.10163) (2048,1.06828) (4096,1.02865) (8192,0.993847) (16384,0.979935) (33768,0.983348) (65536,0.998581) (131072,0.975691) (262144,0.995023) (524288,1.03199) (1048576,0.97128) };
    \addlegendentry{\staggered};
    \addplot[draw=none] coordinates { (32,2) (33,2)  };
    \addlegendentry{\hspace{22pt}\deterdupl\\ \mbox{(\antbams deadlocks)}};

    \addplot[color=black] coordinates { (1,1) (1048576,1) };
  \end{axis}
\end{tikzpicture}
\label{fig:ams robust}
\end{subfigure}
\begin{subfigure}[b]{0.5\textwidth}

\caption{\aams vs. \andaams}
\vspace{-0.2cm}
\begin{tikzpicture}
  \begin{axis}[
    xmin=1,
    xmax=1048576,
    xlabel={$\frac{n}{p}$},
    ylabel style={align=center},
    ylabel={Running Time Ratio\\$[t_{\text{\aams}} / t_{\text{\andaams}}]$},
    legend style={
      at={(0.5,0.97)},anchor=north},
    legend columns=3,
    plotconfig,
    ymax = 2.4,
    ymin=-0.1,
    xmode=log,
    log base x=2,
    y label style={at={(0.04,0.5)}},
    ytick={0, 0.5, 1, 1.5, 2},
    xtickten={0, 5, 10, 15, 20},
    ]
    \addplot coordinates { (1,1.04) (2,1.06524) (4,1.0578) (8,1.06192) (16,1.05646) (32,1.05518) (64,1.15268) (128,1.32483) (256,1.30245) (512,1.19376) (1024,1.10982) (2048,1.0463) (4096,1.02775) (8192,1.02652) (16384,1.01927) (33768,1.01255) (65536,1.01251) (131072,1.00714) (262144,1.00368) (524288,0.990584) (1048576,0.997463) };
    \addlegendentry{\uniform};
    \addplot coordinates { (1,1.04866) (2,1.07038) (4,1.05969) (8,1.06701) (16,1.05747) (32,1.06249) (64,1.06196) (128,1.07183) (256,1.06005) (512,1.05327) (1024,1.05718) (2048,1.05352) (4096,1.04601) (8192,1.03644) (16384,1.02474) (33768,1.00968) (65536,1.00335) (131072,1.01069) (262144,1.00567) (524288,0.995176) (1048576,0.996771) };
    \addlegendentry{\bucketsorted};
    \addplot coordinates { (1,1.04348) (2,1.07119) (4,1.07385) (8,1.07125) (16,1.05522) (32,1.06702) (64,1.06836) (128,1.07088) (256,1.0646) (512,1.06149) (1024,1.05466) (2048,1.04668) (4096,1.03654) (8192,1.02781) (16384,1.02401) (33768,1.00131) (65536,0.997424) (131072,1.01023) (262144,1.02283) (524288,1.02036) (1048576,1.01386) };
    \addlegendentry{\staggered};
    \addplot coordinates { (1,1.04653) (2,1.06932) (4,1.06303) (8,1.07006) (16,1.05825) (32,1.06471) (64,1.06813) (128,1.06262) (256,1.05953) (512,1.04338) (1024,1.05046) (2048,1.05064) (4096,1.04501) (8192,1.02581) (16384,1.01473) (33768,1.0119) (65536,1.00617) (131072,0.995424) (262144,0.999868) (524288,1.00295) (1048576,0.999187) };
    \addlegendentry{\deterdupl};
    \addplot coordinates { (1,1.03862) (2,1.06288) (4,1.06288) (8,1.06672) (16,1.0565) (32,1.06668) (64,1.2458) (128,1.40131) (256,1.17243) (512,0.853731) (1024,0.608908) (2048,0.390853) (4096,0.301998) (8192,0.247514) (16384,0.208779) (33768,0.192099) (65536,0.195849) (131072,0.198282) (262144,0.341006) (524288,0.547142) (1048576,0.760861) };
    \addlegendentry{\alltoone};

    \addplot[color=black] coordinates { (1,1) (1048576,1) };
  \end{axis}
\end{tikzpicture}
\label{fig:determ msg assignment}
\end{subfigure}%
\begin{subfigure}[b]{0.5\textwidth}

\caption{\aams vs. \asamplesort and \andsams}
\vspace{-0.45cm}
\begin{tikzpicture}
  \begin{axis}[
      xmin=1,
      xmax=1048576,
      ymax=2.4,
      ymin=-0.1,
      legend style={
        at={(0.5,0.97)},anchor=north},
      ylabel style={align=center},
      ylabel={$[t_{\text{\aams}} / t_{\text{\andsams}}]$;$[t_{\text{\aams}} / t_{\text{\asamplesort}}]$},
      xlabel={$\frac{n}{p}$},
      plotconfig,
      legend columns=2,
      xmode=log,
      log base x=2,
      y label style={at={(0.04,0.45)}},
      ytick={0, 0.5, 1, 1.5, 2},
      xtickten={0, 5, 10, 15, 20},
    ]
    \addplot coordinates { (1,0.0347374) (2,0.0355076) (4,0.0361572) (8,0.0373108) (16,0.039197) (32,0.0426594) (64,0.0515414) (128,0.0622241) (256,0.0630069) (512,0.0647214) (1024,0.0672186) (2048,0.0705144) (4096,0.0806837) (8192,0.0860181) (16384,0.102771) (32768,0.135664) (65536,0.178682) (131072,0.250781) (262144,0.387435) (524288,0.524091) (1048576,0.647782) };
    \addlegendentry{\andsams};
    \addplot coordinates { (1,0.0269945) (2,0.0240181) (4,0.0197335) (8,0.0141509) (16,0.0093301) (32,0.00557132) (64,0.00351148) (128,0.00209562) (256,0.00102211) (512,0.000971415) (1024,0.00103245) (2048,0.00110256) (4096,0.00126056) (8192,0.00140841) (16384,0.00189143) (32768,0.00297774) (65536,0.00506201) (131072,0.00953623) (262144,0.0196431) (524288,0.0372535) (1048576,0.067431) };
    \addlegendentry{\asamplesort};

    \addplot[color=black] coordinates { (1,1) (1048576,1) };
  \end{axis}
\end{tikzpicture}
\label{fig:simple sample sort}
\end{subfigure}
\vspace{-1cm}
\caption{Running time ratios of robust algorithms to nonrobust algorithms.
\ref{fig:quick robust}) Ratios of \aquick to \antbquick (262\,144 cores).
\ref{fig:ams robust}) Ratios of \aams to \antbams (8\,192 cores).
\ref{fig:determ msg assignment}) Ratios of \aams to \andaams (131\,072 cores).
\ref{fig:simple sample sort}) Ratios of \aams to \andsams (131\,072 cores).}
\label{fig:robust sparse}
\end{figure*}
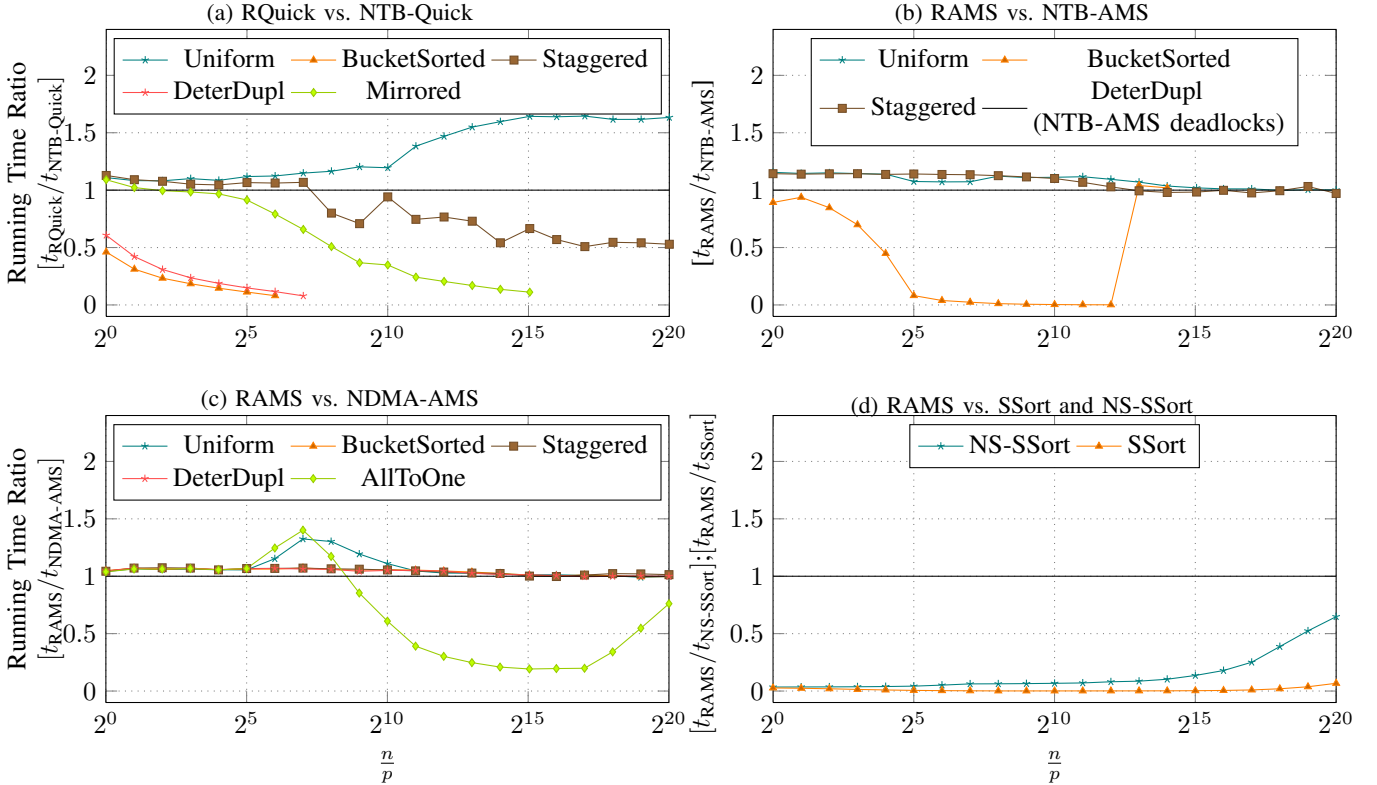

We present the results of a simple binomial-tree gather-merge (\emph{\agather}), all-gather-merge (\emph{\aallgather}) from Section~\ref{s:hypercube architecture}, robust fast work-inefficient sorting (\emph{\afis}) from Section~\ref{s:fastInefficient ams}, robust quicksort on hypercubes (\emph{\aquick}) from Section~\ref{s:quicksort}, robust AMS-sort (\emph{\aams}) from Section~\ref{s:fastInefficient ams}, and a simple \mbox{$p$-way} sample sort implementation (\emph{\asamplesort}).
We compare the results with our closest competitors, HykSort~\cite{Sundar13} (\emph{\ahyk}) and bitonic sort (\emph{\abitonic}) which \ahyk used for small inputs.
We ran our experiment on JUQUEEN, an IBM BlueGene/Q based system with a  $5$-D torus network and spent more than four million~core-hours.

We ran benchmarks with eight input instances proposed by Helman et al.~\cite{Helman98}.
We report results for the input instances  
\emph{\uniform} (independent random values),
\emph{\bucketsorted} (locally random, globally sorted), 
\emph{\deterdupl} (only $\log p$ different keys), and
\emph{\staggered} (both designed to be difficult for hypercube-like routing).
We do not show results for \gaussian, \ggroup, \zero, and \randdupl as their results are more or less represented by results we show:
\uniform is similar to \gaussian;
\bucketsorted is similar to \ggroup and \zero with the difference that \aquick sorts the instances \ggroup and \zero with more than $2^{17}$ elements twice as slow as \aams;
\deterdupl is similar to \randdupl.
We also included the input instances \emph{\mirrored} and \emph{\alltoone}.
The input of \mirrored on PE $i$ are random numbers between $2^{31}(m_i)/p$ and $2^{31}(m_i + 1)/p$ where the bit representation of $m_i$ is the reverse bit representation of $i$.
After $\frac{\log p}{2}$ recursions of a naive implementation of hypercube quicksort, $\ceiling{\sqrt{p}}$ PEs hold $\flooring{n/ \sqrt{p}}$ elements each. 
The first $n/p - 1$ elements of \alltoone  on PE $i$ are random elements from the range $[p + (p - i) (2^{32} - p)/p, p + (p - i + 1) (2^{32} - p)/p]$ and the last element has the value $p - i$. 
At the first level of a naive implementation of $k$-way sample sort, the last $\text{min}(p, n/p)$ PEs would send a message to PE $0$.

In Section~\ref{sec:results largest instance}, we present results for our algorithms and our closest competitors on 262\,144 cores (16\,384 nodes), the largest available jobs on JUQUEEN.
In Section~\ref{sec:robustness analysis}, we compare our robust algorithms to their nonrobust versions.
We show that the nonrobust counterparts are orders of magnitude slower for hard inputs.
See Appendix~\ref{app:experiments} for more information on JUQUEEN, the input instances, and the benchmarking details.
We also give implementation details in Appendix~\ref{app:implementation}, describe our extensive parameter tuning on 132\,072 cores in Appendix~\ref{app:parameter tuning}, and compare \aams and \ahyk to the literature in Appendix~\ref{app:comparison to the literature}.

\subsection{Input Size Analysis and Algorithm Comparison}\label{sec:results largest instance}

In this section we present the results of our experiments on all algorithms executed on 262\,144 cores.
The smallest inputs have only one element every $3^5$ (243) PEs whereas the largest input has $2^{23}$ elements per core.

Figure~\ref{fig:largest instance rt} shows the running times of each algorithm for the most interesting input instances (Appendix~\ref{app:largest instance ratio} gives running time ratios of each algorithm compared to the fastest algorithm for the most interesting input instances).
We now explain the most important results.
(1) \agather sorts very sparse inputs ($n/p \leq 3^{-3}$) up to $1.8$ times faster than the other sorting algorithms.
\aallgather is not competitive for any input size and sorts even the sparsest input twice as slow as \afis.
However, neither fulfills the balance constraint of sorted output.
(3) For all input instances, \afis is the fastest sorting algorithm when $3^{-3}<n/p\leq 4$.
For example, \afis sorts \uniform sparse inputs up to $3.6$ times faster than its competitors and a single element per core more than twice as fast as \aquick and \abitonic.
(4) \aquick sorts small inputs ($2^3$ to $2^{14}$) of \uniform instances up to $3.4$ times faster than any other algorithm (more than 5 times faster if we exclude our own algorithms).
In general, the running times of \aquick for other input instances do not differ much for sparse and small inputs.
For $2^{15}$ and $2^{16}$ elements per PE, \aams is the fastest algorithm.
For $n/p \geq 2^{17}$, the situation becomes more instance dependent.
For uniform inputs, \ahyk is 1.22 times faster than \aams for the largest inputs (and 1.38 times faster for the isolated case $n/p=2^{19}$). However,  \ahyk crashed on input instances \deterdupl, \randdupl, \bucketsorted, \ggroup, and \zero and is up to 1.7 times slower than \aams for \staggered inputs. Overall, \aams seems to be a good compromise between robustness and performance for large inputs. \aams is up to 2.7 (\bucketsorted) times faster than \aquick.

\subsection{Robustness Analysis}\label{sec:robustness analysis}

We now have a closer look at the impact of various algorithmic measures to improve robustness.

\textbf{Robust Quicksort:} Figure~\ref{fig:quick robust} depicts the running time ratio of \aquick over \aquick   without redistribution (\antbquick).
The price of robustness for simple input instances such as \uniform is an additional data redistribution.
For large \uniform inputs, this slows down \aquick by a factor of up to $1.7$.
On the other hand, \aquick sorts skewed input instances such as \staggered and \mirrored robustly.
Thus, the additional redistribution of \aquick decreases the total running time for large inputs  by a factor of up to $9$ ($n = 2^{15}$) and \aquick runs out of memory for $n = 2^{16}$.
Our results show that \aquick sorts other input instances (\bucketsorted and \deterdupl) faster than \antbquick by several orders of magnitude due to tie-breaking.
\antbquick even fails to sort such input instances when the input is large.

\textbf{Tie-Breaking of \aams:}
Figure~\ref{fig:ams robust} shows the running time ratio of \aams and \aams without tie-breaking during local data partitioning (\emph{\antbams}).
When sorting small inputs without unique keys (e.g., \uniform and \staggered), the extra work to calculate splitters with tie-breaks slows down \aams compared to \antbams (factor of $1.15$ for $n/p < 2^{12}$). However, this effect becomes negligible for the large inputs ($n/p\geq 2^{16}$) \aams is intended for (i.e., when it outperforms \aquick). 
\antbams sorts small inputs of the instance \bucketsorted much slower than \aams as small instances of \bucketsorted input contain many duplicates.  The non-robust version deadlocks immediately when we sort \deterdupl and \zero instances.

\textbf{Message Assignment of \aams:}
\frage{heavily reformulated because the discussion required detailed knowledge of the SPAA paper.}Figure~\ref{fig:determ msg assignment} shows the running time ratio of \aams using $l=3$ levels of data redistribution and deterministic message assignment (\adma) on 131\,072 cores with the same algorithm without \adma (\andaams).
\aams decides to sort \staggered, \bucketsorted, and \deterdupl inputs without \adma as there would be no impact of \adma. Our experiments show that the overhead for making that decision is small.
\aams actively performs \adma for small inputs ($64$ to $2^{10}$) of \uniform instances leading to a certain overhead. However, for such small instances \aquick would be used preferable in any case.
For the input instance \alltoone, \andaams sends  $\Oh{\text{min}(n/p, p)}$ messages to PE $0$ in the first level.
In this case, \aams actually can take advantage of \adma as it reduces the startups to $\Oh{k}$.
This speeds up \aams by a factor of up to $5.2$.
In Figure~\ref{fig:determ msg assignment}, the positive effect begins for $n/p > 8 k = 2^9$ elements per core and increases rapidly.
The effect decreases when the time for the message exchange dominates the startup time.

\textbf{Simple Sample Sort:} The comparisons above are for recent, highly sophisticated massively parallel sorting algorithms. To give an impression how big the improvements of the new algorithms are compared to the majority of parallel sorting algorithms presented in the past for moderate number of PEs, we also compare 3-level \aams with a simple $p$-way sample sort that delivers the data directly. 
Figure~\ref{fig:simple sample sort} shows that
\aams for \uniform instances is up to 1\,000 times faster than \asamplesort. \aams is much faster even if we ignore the time for finding splitters (\emph{\andsams}).
When we look at the range where \aams is faster than \aquick (at least $2^{15}$ elements per core), \aams outperforms \andsams by a factor of $1.5$ to $7.4$.
Experiments show that this effect increases as $p$ increases.
Note that the curve for \andsams acts as a rough lower bound for any algorithm that delivers the data directly.

\section{Conclusion and Future Work}

We show how to obtain robustness of massively parallel sorting with respect to the input size by using three different algorithms
(\afis, \aquick, and \aams). Robustness with respect to skew can be achieved by careful randomization (\aquick) and message assignments (\aams). Robustness with respect to duplicate keys can be achieved at reasonable overhead using careful implicit implementations of the brute-force tie-breaking idea. We plan to release an open-source library with our algorithms.

Still further improvements are possible.
We could remove the limitation to $p$ being a power of two. This will complicate the code but should not be a fundamental obstacle. For large inputs, a specialized shared-memory implementation for node-local sorting, merging and partitioning seems useful. Porting to systems with less deterministic behavior than BlueGene/Q is likely to require sophisticated implementations of collective communication and data delivery operations -- the wide fluctuations of running times mentioned in \cite{Axtmann15} are already quite bad on $2^{15}$ PEs on an oversubscribed InfiniBand-Network.

\emph{Acknowledgement.} In Appendix~\ref{app:ack} we acknowledge the Gauss Centre for Supercomputing (GCS) for providing computation time, the DFG for partially funding this research and the SAP AG, Ingo M\"uller, Sebastian Schlag, Helman et al.~\cite{Helman98}, and Sundar et al.~\cite{Sundar13} for making their code available.

\balance  
\bibliographystyle{IEEEtran}

\bibliography{diss}

\clearpage

\appendix

\subsection{Model of Computation}\label{app:model of computation}

A common abstraction of communication in supercomputers is the (symmetric) single-ported message passing model.
It takes time $\alpha + l\beta$ to send a message of size $l$ machine words.
The parameter $\alpha$ defines the startup overhead of the communication.
The parameter $\beta$ defines the  time to communicate a machine word.
For simplicity, we assume that the size of a machine word is equivalent to the size of a data element.
For example, broadcast, reduction, and prefix sums can be implemented to run in time $\Oh{\beta l + \alpha \log p}$~\cite{Batcher68, SST09} for vectors of size $l$. We have $\alpha\gg\beta\gg 1$
where our unit is the time for executing a simple machine instruction. Most of the time, we treat $\alpha$ and $\beta$ as variables in our asymptotic analysis in order to expose effects of latency and bandwidth limitations.

\subsection{Hypercube Algorithms}
\label{app:hypercube architecture}
\begin{algorithm}
\begin{algorithmic}
  \caption{Hypercube algorithm design pattern}\label{algo:app:hypercube}
  \State Computation on PE $i$ 
  \For{$j\Is d - 1$ {\bf downto} 0} \Comment{or $0..d-1$}
    \State send some message $m$ to PE $i\oplus 2^j$
    \State receive message $m'$ from PE $i\oplus 2^j$
    \State perform some local computation using $m$ and $m'$
  \EndFor
\end{algorithmic}
\end{algorithm}
A hypercube network of dimension $d$ consists of $p=2^d$ PEs numbered $\set{0,\ldots,p-1}$%
.
Two nodes $a$ and $b$ are connected along dimension $i$ if $a=b\oplus 2^i$.
For this paper, hypercubes are not primarily important as an actual network architecture. Rather, we extensively use communication in a conceptual hypercube as a design pattern for algorithms.
More specifically the hypercube Algorithms~\ref{algo:app:hypercube} iterate through the dimension of the hypercube. 
Depending on how this template is instantiated, one achieves a large spectrum of global effects. 
For example, by repeatedly summing an initial local value $a$, one gets an \emph{all-reduce}, i.e., all PEs get the global sum of the local values in time $\Oh{(\alpha+\beta |a|)\log p}$. 
Similarly, if we replace addition by concatenation, we perform an \emph{all-gather} operation (i.e., all PEs get all the local values) which runs in time 
$\Oh{\beta p |a| + \alpha\log p}$. 
If the $a$s are a sorted sequences and we replace concatenation by merging,
all PEs get the elements in all the local $a$s in sorted order using time $\Oh{\beta p |a| + \alpha\log p}$. We call this operation \emph{all-gather-merge}.

We can also use a hypercube algorithm for routing data -- in iteration $j$, a data object destined for PE $t$ and currently located on PE $i$ is moved if $t$ and $i$ differ in bit $j$ (e.g. \cite{Lei92}). This has the advantage that we need only $\Oh{\log p}$ startup overheads overall. However, for worst case inputs, even if every PE sends and receives only a single object, 
the required time can be $\Om{\sqrt{p}}$ since many objects can visit the same intermediate nodes. However, it is known that for random start or destination nodes, the running time remains $\Oh{\log p}$ \cite{Lei92}.

To describe and understand hypercube algorithms, we need the concept of a subcube. A $j$-dimensional subcube consists of those PEs whose numbers have the same bits $j..d-1$ in their binary representation.

\subsection{Randomized Shuffling on Hypercubes}\label{app:randomShuffle}

It is a folklore observation (e.g. \cite{sanders97}) that data skew can be removed by shuffling the input data randomly. This is actually implemented for the sample sort algorithm of Helman et al.~\cite{Helman98} by directly sending each element to a random destination. Note that this incurs an overhead of about 
$\alpha p+\beta n/p$. We propose to use this technique for small inputs where we need smaller latency. This can be achieved by routing the data according to a hypercube communication pattern -- sending each element to a random side in each communication step. To achieve even slightly better load balance, we actually split local data in two random halves in each communication step. The resulting running time is
\begin{align*}
\Oh{\left(\alpha+\beta \frac{n}{p}\right)\log p}\punkt
\end{align*}
A naive approach labels each element with a random destination. Then, a hypercube algorithm redistributes the data. This approach increases the communication volume by a factor of two.

\subsection{Sorting Algorithms from Tiny to Huge Inputs}
\label{app:related work}

\subsubsection{Fast Work-Inefficient Ranking}
\label{app:fast work ranking}

We recently proposed a simple but fast ranking algorithm for small inputs~\cite{Axtmann15}.
The PEs are arranged in an array of size $\Oh{\sqrt{p}} \times \Oh{\sqrt{p}}$ and each PE has $\Oh{\frac{n}{p}}$ elements.
We first sort the elements locally in $\Oh{\frac{n}{p} \log \frac{n}{p}}$.
Then we all-gather-merge the elements of PEs in the same row as well as between PEs in the same column in 
$\Oh{\alpha \log p + \beta n/\sqrt{p}}$.
Afterwards, each PE stores the elements of all PEs in its row and column respectively in a sorted way.
Then the PEs rank each element received from their column in elements received from their row.
Finally, we sum up the local ranks in each column with an allreduce operation. The result is that each PE knows the global rank of all inputs elements in its row. In theory this is a very good algorithm for $n=\Oh{\sqrt{p}}$ since we get only logarithmic delay.
However, since all other algorithms need $\Om{\log^2p}$ startup overheads, Fast work-inefficient ranking is interesting up to $n=\Oh{\frac{\alpha}{\beta}\sqrt{p}\log^2p}$. In Subsection~\ref{app:fastInefficient} we make the algorithm robust against identical keys and show how to convert its output to a classical sorted permutation of the input.

\subsubsection{Bitonic Sort}

Bitonic sort~\cite{Batcher68,johnsson84} first sorts locally and then performs $\Oh{\log^2p}$ pairwise exchange and merge operations.

For small inputs this running
time is dominated by $\log^2p$ startup overheads.  This gives it a similar running time as the parallel quicksort algorithms to be
discussed next.  However, for $n=\omega(p\alpha/\beta)$ the term
$\beta \frac{n}{p} \log^2 p$ dominates -- all the data is exchanged $\log^2
n$ times. This makes it unattractive for large inputs compared to
quicksort and other algorithms designed for large inputs. Indeed, only
for $n$ superpolynomial in $p$ ($n=\Omega(p^{\log p})$), bitonic sort eventually becomes
efficient. But for such large inputs algorithms like sample sort are much
better since they exchange the data only once.
\subsubsection{Parallel Quicksort}

Since quicksort is one of the most popular sorting algorithms, it is not surprising that there are also many parallel variants. For distributed memory, a variant using the hypercube communication pattern is attractive since it is simple and can exploit locality in the network. A recursive subproblem is solved by a subcube of the hypercube. The splitter is broadcast to all PEs. Elements smaller than the pivot are moved to the 0-subcube and elements larger than the pivot are moved to the right 1-subcube.
Since this hypercube quicksort obliviously splits the PEs in half, the imbalance accumulating over all the levels of recursion is a crucial problem. In its simplest original form by Wagar \cite{wagar87},
PE 0 uses its local median as a pivot. This is certainly not robust against skew and even for average case inputs it only works for rather large inputs. One can make this more robust -- even in a deterministic sense -- by using the global median of the local medians \cite{Lan92}. 
However, this introduces an additional $\beta p$ term into the communication complexity and thus defeats the objective of having polylogarithmic execution time. Therefore in Subsection~\ref{s:quicksort}
we propose to use a median selection algorithm that is both fast and accurate. 

Non-hypercube distributed memory quicksort is also an interesting option where we can adapt the number of PEs working on a recursive subproblem to its size. However, it leads to more irregular communication patterns and has its own load balancing problems because we cannot divide PEs fractionally.
Siebert and Wolf \cite{SieWol11} exploit the special case where this is no problem, when $n=p$.

\subsubsection{Generalizing Quicksort}

For large inputs it is a disadvantage that quicksort exchanges the data at least $\log p$ times. We can improve that by partitioning the input with respect to $k-1$ pivots at once. This decreases the number of times data is moved to $\Oh{\log_kp}$. However, the price we pay is 
latency $\Om{\alpha k}$ for delivering data from $k$ partitions to $k$ different PEs.
This gives us a lower bound of 
$$\OmL{\frac{n}{p}\log n}+\beta\frac{n}{p}\log_kp+\alpha k\log_kp$$ 
for running generalized quicksort. Getting an algorithm with a matching upper bound is not easy though. We have to find pivots that partition the input in a balanced way and we have to execute the more complex data delivery problems efficiently.

Many algorithms have been devised for the special case $k=p$.
Sample sort \cite{BleEtAl91} is perhaps most well known because it is simple and effective. It achieves the above bound if a sample of size $S=\Om{p\log n}$ is sorted using a parallel algorithm. Then, every $S/k$-th sample is used as a splitter. This algorithm achieves the desired bound for $n=\Oh{p^2/\log p}$.
Solomonik and Kale \cite{SolKal10} describe the best scaling single level algorithm we have seen and scales to $2^{15}$ PEs for large inputs.
One can even achieve perfect partitioning by finding optimal splitters in a multiway mergesort approach \cite{VarEtAl91,SSP07,Axtmann15}. This requires $n=\Oh{p^2\log p}$ for achieving the desired bound.

Gerbessiotis and Valiant \cite{GerVal94} develop a multilevel sample sort for the BSP model \cite{Val94}. However, implementing the data exchange primitive of BSP requires $p$ startup overheads in each round. See \cite{Axtmann15} for a more detailed discussion of previous theoretical multilevel algorithms. 

Sundar et al. develop and implement \emph{HykSort}, a generalized hypercube quicksort \cite{Sundar13}. HykSort uses a sophisticated recursive sampling based algorithm for finding high quality approximate splitters that is something like a compromise between the single shot algorithms used in sample sort and the exact (but slower) algorithms used for multiway mergesort. This makes HykSort ``almost'' robust against skew. However, in the worst case, data may become severely imbalanced.
The paper mentions measures for solving this problem, but does not analyze them.
Furthermore, HykSort uses the operations {\tt MPI\_Comm\_Split} whose current implementations need time $\Om{\beta p}$. This is why we put a ``$\geq$'' in the corresponding row of Table~\ref{t:time}. HykSort is also not robust with respect to duplicate keys.

\emph{AMS-sort}~\cite{Axtmann15} our recently published $k$-way sample sort algorithm which guarantees a maximum imbalance of $\epsilon \frac{n}{p}$ for arbitrary inputs. Within each recursion, AMS-sort sorts $\Oh{k \log k}$ samples in parallel. Then it selects and distributes $b \cdot k$ splitters where $b = \Oh{\frac{1}{\sqrt{1 + \epsilon} - 1}}$. After the local data has been partitioned into $b \cdot k$ partitions, a greedy algorithm assigns global partitions to PE subcubes by minimizing the load imbalance between subcubes. Then each PE calculates for its partitions the target PEs within the subcubes. 
This ensures perfect load balancing within target subcubes. 
Thus, by going away from a (even generalized) hypercube communication pattern, AMS-sort avoids the data imbalance that may make HykSort inefficient. Our paper~\cite{Axtmann15} describes (but does not implement) a way to select communication partners such that each PE sends and receives $\Theta(k)$ messages for a $k$-way data exchange. 
Similarly, a tie-breaking technique is proposed (but not implemented) which simulates unique keys with very low overhead.
In Subsection~\ref{app:ams} we describe our robust implementation of AMS-sort.

\subsection{More Related Work}
\label{app:more related work}

This paper was highly influenced by two landmark papers. 25 years ago, Blelloch et al. \cite{BleEtAl91} carefully studied a large number of algorithms, selected three of them (bitonic, radix, and sample sort),
and compared them experimentally on up to $2^{15}$ PEs. Most other papers are much more focused on one or two (more or less) new algorithms sometimes missing the big picture. We felt that it was time again for a more wide view in particular because it became evident to us that a single algorithm is not enough on the largest machine and varying $n/p$.
Helman et al. \cite{Helman98} -- concentrating on sample sort -- took robustness seriously and made systematic experiments with a wide variety on input distributions which form the basis for our experiments. They also propose initial random shuffling to make samplesort robust against skewed input distributions (this was already proposed in \cite{SanHan97e}).

\subsection{Robust Fast Work-Inefficient Sorting}\label{app:fastInefficient}

We propose a fast sorting algorithm (\afis) for sparse and very small inputs
that is robust against duplicate keys.
Our algorithm computes the rank of each element in the global data.
Then, optionally, a data delivery routine sends the elements to the appropriate PEs according to their ranks.
For ranking the input data, we use the algorithm we recently proposed~\cite{Axtmann15} (see also~\cite{IKS09} and~\ref{app:fast work ranking}).
Unfortunately, the proposed approach calculates the same rank for equal elements.
However, our data delivery techniques require unique ranks in the range of $0..n-1$ as input to achieve balanced output.

\begin{figure}\normalsize\centering
  \includegraphics{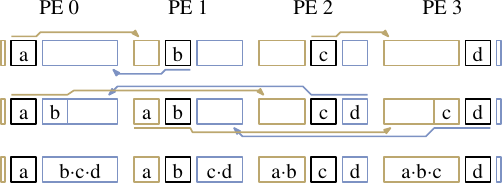}
  \caption{All-gather-merge keeps track where elements came from. Each PE has three buckets; one bucket stores elements which came from the left (brown), one bucket stores elements which came from the right (blue), and one bucket stores elements initially provided by this PE (black).}
  \label{fig:fast inefficient}
\end{figure}

Consider a tie-breaking policy that logically treats
an element as a quadruple $(x,r,c,i)$ where $x$ is the key, 
$r$ the row, $c$ the column, and $i$ the position in the locally sorted data.
We can then define a total ordering between these quadruples as their lexicographical order. 
A modified all-gather-merge and a modified compare function implement that policy without communication overhead for communicating the $(r,c,i)$ information.
For the all-gather-merge in a row, each PE uses three arrays $\set{\leftarrow, H,  \rightarrow}$ to keep track which elements came from the left ($\leftarrow)$, which elements came from the right ($\rightarrow$) and which elements are local elements ($H$).
Assume that we implement all-gather-merge with a hypercube algorithm starting from the lowest dimension.
When a PE receives data from left (right), \emph{all} elements in that message come from left (right) and are merged into the array $\leftarrow$ ($\rightarrow$).
Analogously, for the all-gather-merge in a column, PEs merge elements from above into the array $\uparrow$ and elements from below into the array $\downarrow$.
This classification suffices to perform tie-breaking according to the lexicographic ordering of the quadruples.
Figure~\ref{fig:fast inefficient} shows the all-gather-merge algorithm in a row.
We now update the compare function on PE $(r,c)$ between an element $a=(x,r,C,i)$ from row $r$ and an element  $b=(y,R,c,j)$ from column $c$ where the arrays $C\in\set{\leftarrow,H,\rightarrow}$ and $R\in\set{\uparrow,H,\downarrow}$ are used as implicit element labels.
Furthermore, $i$ ($j$) specifies the position among the locally sorted data if $C=c$ ($R=r$) and is undefined otherwise.
Then the following table defines the updated compare function of comparing $a > b$
\begin{center}
\begin{tabular}{c| >{\centering}m{2cm} >{\centering}m{2cm} >{\centering}m{2cm}}
$b\setminus a$ & $\leftarrow$ & $c$    & $\rightarrow$ \tabularnewline\hline
$\uparrow$     & $a \geq b$       & $a \geq b$ & $a \geq b$        \tabularnewline
  $r$            & $a > b$       & $i> j$ & $a \geq b$           \tabularnewline
$\downarrow$   & $a > b$          & $a > b$    & $a > b$
\end{tabular}
\end{center}
When we rank row elements into column elements, we choose the correct compare function just once for each combination of row array $\set{\leftarrow, H, \rightarrow}$ and column array $\set{\uparrow, H, \downarrow}$.

After we computed the ranks of the input elements, we redistribute them to get sorted output.

\subsection{Robust Multi-Level Sample Sort}
\label{app:ams}

The prototypical implementation of AMS in \cite{Axtmann15} does not ensure that each PE communicates with $\Oh{k}$ PEs in each
level of recursion. Similarly, \cite{Axtmann15} proposes but does not implement 
a tie-breaking scheme to become robust against duplicate keys.
Here, we present a complete implementation of AMS-sort (\aams) including both improvements outlined above.

Even for worst case inputs, the following partitioning approach splits input as balanced as for input with unique keys.
To do so, we first pick random samples of the local data and add their positions in the input as tie-breakers to the samples.
Then we rank the samples with the fast work-inefficient ranking algorithm (see Appendix~\ref{app:fast work ranking}).
Next, we select splitters from the samples and distribute them to all PEs.
Finally, the partitioner of Super Scalar Sample Sort~\cite{SW04} assigns all local elements to a partition.
We modify the element classifier of Super Scalar Sample Sort to support tie-breaking.
Initially, we classify each input element using its original key.
If this key is equal to the original key of the bounding splitter, we
repeat the search using the positions of the splitters as tie-breakers.
To become cache efficient, we perform multiple partitioning passes if the number of partitions is large.

Our robust version of AMS-sort implements the \emph{deterministic message assignment} proposed for AMS-sort.
The message assignment algorithm first routes information about messages to groups of PEs where the actual message belongs.
Then each group calculates for each message one or more addresses.
Finally, these addresses are sent back to the PEs which will send the actual messages back to the PE group.
However, the PEs do not know the number of assigned addresses and thus do not know the number of messages they have to receive.
The algorithm NBX~\cite{Hoefler2010art}, a dynamic sparse data exchange technique, uses an non-blocking barrier to identify when the message exchange is complete.
We use this method to efficiently exchange the addressees in $\Oh{\alpha \log p + \alpha k}$ time.

\subsection{Experimental Comparison of Median Approximation Trees}
\label{app:median selection trees experiments}

\pgfplotscreateplotcyclelist{my exotic}{%
teal,every mark/.append style={fill=teal!80!black},mark=star\\%
orange,every mark/.append style={fill=orange!80!black},mark=triangle*\\%
brown!60!black,every mark/.append style={fill=brown!80!black},mark=square*\\%
red!70!white,mark=star\\%
lime!80!black,every mark/.append style={fill=lime},mark=diamond*\\%
red,densely dashed,every mark/.append style={solid,fill=red!80!black},mark=*\\%
yellow!60!black,densely dashed,every mark/.append style={solid,fill=yellow!80!black},mark=square*\\%
black,every mark/.append style={solid,fill=gray},mark=otimes*\\%
blue,densely dashed,mark=star,every mark/.append style=solid\\%
red,densely dashed,every mark/.append style={solid,fill=red!80!black},mark=diamond*\\%
}

\pgfplotsset{
  width=\textwidth,
  height=54mm,
  major grid style={thin,dotted,color=black!50},
  minor grid style={thin,dotted,color=black!50},
  grid,
  every axis/.append style={
    line width=0.5pt,
    tick style={
      line cap=round,
      thin,
      major tick length=4pt,
      minor tick length=2pt,
    },
  },
  legend cell align=left,
  legend pos=north west,
}

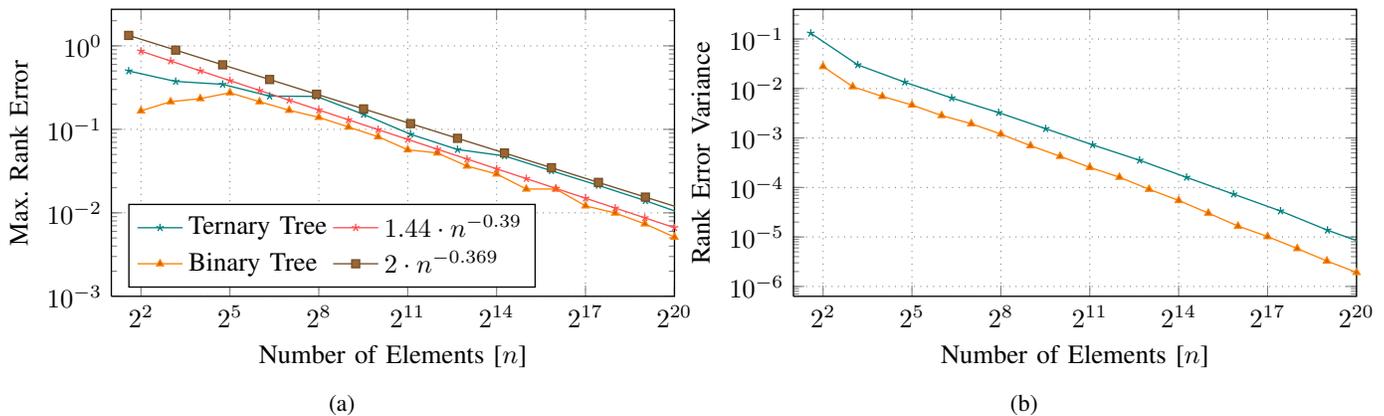
\begin{figure*}
\begin{subfigure}[b]{0.5\textwidth}
        \begin{tikzpicture}
      \begin{axis}[
          xlabel={Number of Elements [$n$]},
          ylabel={Max. Rank Error},
          xmode=log,
          xmin=2,
          xmax=1048576,
          ymin=0.001,
          log basis x = 2,
          ymode=log,
          log basis y = 10,
          legend columns=3,
          legend pos=south west,
          transpose legend,
          mark size=1.5pt,
          cycle list name=my exotic
        ]
        \addplot coordinates { (3,0.5) (9,0.375) (27,0.346154) (81,0.25) (243,0.247934) (729,0.151099) (2187,0.087374) (6561,0.057317) (19683,0.048064) (59049,0.031889) (177147,0.021304) (531441,0.013979) (1594323,0.00888) };
        \addlegendentry{Ternary Tree};
        
        \addplot coordinates { (4,0.166667) (8,0.214286) (16,0.233333) (32,0.274194) (64,0.214286) (128,0.169291) (256,0.139216) (512,0.106654) (1024,0.081623) (2048,0.056913) (4096,0.052137) (8192,0.03632) (16384,0.029207) (32768,0.019272) (65536,0.01928) (131072,0.012112) (262144,0.009951) (524288,0.007379) (1048576,0.005122) (2097152,0.003919) (4194304,0.00295) };
        \addlegendentry{Binary Tree};

        \addplot coordinates { (3,1.33344) (9,0.889026) (27,0.59273) (81,0.395184) (243,0.263476) (729,0.175664) (2187,0.117119) (6561,0.0780851) (19683,0.0520607) (59049,0.0347098) (177147,0.0231417) (531441,0.015429) (1594323,0.0102868) };
        \addlegendentry{$2\cdot n^{-0.369}$};

        \addplot coordinates { (4,0.862766) (8,0.658402) (16,0.502446) (32,0.383431) (64,0.292608) (128,0.223298) (256,0.170405) (512,0.130041) (1024,0.0992383) (2048,0.0757317) (4096,0.0577931) (8192,0.0441036) (16384,0.0336568) (32768,0.0256845) (65536,0.0196006) (131072,0.0149578) (262144,0.0114147) (524288,0.00871092) (1048576,0.00664756) (2097152,0.00507295) (4194304,0.00387132) };
        \addlegendentry{$1.44\cdot n^{-0.39}$};
      \end{axis}
    \end{tikzpicture}
    \caption{}
  \label{sfig:max rank error}
\end{subfigure}%
\begin{subfigure}[b]{0.5\textwidth}
    \begin{tikzpicture}
      \begin{axis}[
          xlabel={Number of Elements [$n$]},
          ylabel={Rank Error Variance},
          xmin=2,
          xmax=1048576,
          xmode=log,
          log basis x = 2,
          ymode=log,
          log basis y = 10,
          legend columns=3,
          legend pos=south west,
          transpose legend,
          mark size=1.5pt,
          cycle list name=my exotic
        ]
        \addplot coordinates { (3,0.130748) (9,0.0300493) (27,0.0134186) (81,0.0064262) (243,0.00325197) (729,0.00154028) (2187,0.000726576) (6561,0.000354408) (19683,0.000160034) (59049,7.33546e-05) (177147,3.34747e-05) (531441,1.36964e-05) (1594323,6.25831e-06) };
        
        \addplot coordinates { (4,0.0277848) (8,0.0108803) (16,0.00692616) (32,0.00463538) (64,0.00284104) (128,0.00194389) (256,0.00119932) (512,0.000698674) (1024,0.000426226) (2048,0.00025491) (4096,0.000161191) (8192,9.20109e-05) (16384,5.4594e-05) (32768,3.03197e-05) (65536,1.65796e-05) (131072,1.01693e-05) (262144,5.83237e-06) (524288,3.26541e-06) (1048576,1.92179e-06) (2097152,1.13378e-06) (4194304,6.2684e-07) };
      \end{axis}
    \end{tikzpicture}
    \caption{}
  \label{sfig:variance rank error}
\end{subfigure}
    \caption{Maximal rank error (a) and variance of the rank error (b) of the ternary-tree median approximation and the binary-tree median approximation.}
    \label{fig:rank error}
\end{figure*}

We experimentally evaluate the median approximations of the binary-tree selection proposed in Subsection~\ref{s:approxRankSelection} and the ternary-tree selection  proposed by Dean et al.~\cite{dean2014}.
We show that the binary-tree selection gives better median approximations than the ternary-tree selection and the rank can also be bounded by $n/2(1\pm cn^{-\gamma})$ with better constants $c$ and $\gamma$.
Our benchmark runs both selection algorithms on uniformly distributed random integers in the range $[0, 2^{32}-1]$ of up to $2^{20}$ elements.
The input into the binary-tree selection is a power of two whereas the input into the ternary-tree selection is a power of three.
We execute both algorithms 2\,000 times for each input size.
We measure the rank $r$ for each output element and calculate the rank error
\begin{align*}
\abs{\frac{r}{n - 1} - \frac{1}{2}}\punkt
\end{align*}
Figure~\ref{sfig:max rank error} shows the rank error in worst cast.
Compared to the ternary-tree selection, the maximal rank error over 2\,000 runs of the binary-tree selection is smaller for all input sizes. 
Moreover the maximal rank error of the binary-tree selection can tightly bounded by $1.44n^{-0.39}$ but the maximal rank error of the ternary-tree selection is just limited by $2n^{-0.369}$.
Figure~\ref{sfig:variance rank error} depicts the variance of the rank error for the binary-tree selection and the ternary-tree selection.
For small input sizes, the variance of the binary-tree median approximation is smaller by a factor of two.
For large inputs, the difference in the variance even increase to a factor of three.

\subsection{Experimental Results}\label{app:experiments}

We ran our experiment on JUQUEEN, an IBM BlueGene/Q based system of $56$ midplanes, each with $512$ compute nodes.
The maximum number of midplanes available to us was $32$.
Each compute node has one IBM PowerPC A2 $16$-core processor with a nominal frequency of $1.6$ GHz and $16$ GByte of memory.
A $5$-D torus network with $40$~GB/s and $2.5$~\textmu sec worst case latency connects the nodes.
The user can specify the number of midplanes per torus dimension.
In our experiments, we configured the tori such that the size of the dimensions are as equal as possible.

We ran benchmarks with nine input instances of $64$-bit floating point elements.
We use the input instances proposed by Helman et al.~\cite{Helman98}:
\emph{\uniform} (independent random values),
\emph{\gaussian} (independent random values with Gaussian distribution),
\emph{\bucketsorted} (locally random, globally sorted), 
\emph{\deterdupl} (only $\log p$ different keys), 
\emph{\randdupl} ($32$ local buckets of random size, each filled with an arbitrary value from $0..31$),
\emph{\zero} (all elements are equal),
\emph{\ggroup} with $g \coloneqq \sqrt{p}$, and
\emph{\staggered} (both designed to be difficult for hypercube-like routing).
We also included the input instances
\emph{\reverse} (reverse sorted input) and \emph{\alltoone}
where the first $n/p - 1$ elements on PE $i$ are random elements from the range $[p + (p - i) (2^{32} - p)/p, p + (p - i + 1) (2^{32} - p)/p]$ and the last element has the value $p - i$. 
At the first level of a naive implementation of $k$-way sample sort, the last $\text{min}(p, n/p)$ PEs would send a message to PE $0$.

We repeat each measurement six times, ignore the first run (used as a warmup) and average over the remaining five runs. Overall, we spent more than four million core-hours.
We do not show error bars since the ratio between the maximum execution time and the average execution time is consistently less than a few percent. The only exception is all-gather-merge which sometimes fluctuates by up to 50 \% for $n=p$.%
\footnote{An interesting observation is that the deterministic algorithm \abitonic has only negligible fluctuations.
This indicates that the fluctuations introduced by the machine itself are very small.}
The time for building MPI communicators (which can be considerable) is not
included in the running time since this can be viewed as a precomputation that
can be reused over many runs with arbitrary inputs.

\subsubsection{Implementation Details}\label{app:implementation}

\pgfplotscreateplotcyclelist{my exotic}{%
teal,every mark/.append style={fill=teal!80!black},mark=star\\%
orange,every mark/.append style={fill=orange!80!black},mark=triangle*\\%
brown!60!black,every mark/.append style={fill=brown!80!black},mark=square*\\%
red!70!white,mark=star\\%
lime!80!black,every mark/.append style={fill=lime},mark=diamond*\\%
red,densely dashed,every mark/.append style={solid,fill=red!80!black},mark=*\\%
yellow!60!black,densely dashed,every mark/.append style={solid,fill=yellow!80!black},mark=square*\\%
black,every mark/.append style={solid,fill=gray},mark=otimes*\\%
blue,densely dashed,mark=star,every mark/.append style=solid\\%
red,densely dashed,every mark/.append style={solid,fill=red!80!black},mark=diamond*\\%
}

\def\RunnintimeLegend{%
  \legend{
    \aams,
    \abitonic,
    \afis,
    \agather,
    \aallgather,
    \aquick,
    \ahyk
  }%
}

\begin{figure*}[t]\centering
\pgfplotsset{
  plotconfig/.style={
    mark size=1.5pt,
    cycle list name=my exotic,
    height=54mm,
    width=\textwidth,
    xtickten={-5, 0, 5, 10, 15, 20},
    legend style={
      fill=none,
      draw=none},
  }
}
\captionsetup[subfigure]{labelformat=empty}
\begin{subfigure}[b]{0.5\textwidth}

\begin{tikzpicture}
  \begin{axis}[
    xmax=8388608,
    xmin=0.0041,
    ylabel={Running Time Ratio $t/t_{\text{best}}$},
    plotconfig,
    xlabel near ticks,
    legend style={at={(0,1.1)}, anchor=west},
    legend columns=7, 
    log base x=2,
    xmode=log,
    line width=0.5pt,
    tick style={
      line cap=round,
      thin,
      major tick length=4pt,
      minor tick length=2pt,
    },
  major grid style={thin,dotted,color=black!50},
  minor grid style={thin,dotted,color=black!50},
  grid,
  font=\small,
  ymax=6.5,
  ytick={0, 1, 2, 3, 4, 5, 6},
  compat=1.3,
  y label style={at={(-0.08,0.5)}},]
    ]
    \addplot coordinates { (1,11.705) (2,8.48755) (4,6.32837) (8,4.88082) (16,4.88128) (32,4.89089) (64,5.15048) (128,6.14779) (256,5.73469) (512,5.09682) (1024,2.89201) (2048,2.95207) (4096,2.00156) (8192,1.62769) (16384,1.0827) (33768,1) (65536,1) (131072,1) (262144,1.0875) (524288,1.37613) (1.04858e+06,1.2215) (2.09715e+06,1.22177) (4.1943e+06,1.21829) (8.38861e+06,1.22055) };
    \addlegendentry{algo=AMS};
    \addplot coordinates { (1,2.64459) (2,1.93332) (4,1.43581) (8,1.14813) (16,1.55523) (32,1.57033) (64,1.67443) (128,1.93069) (256,2.44781) (512,3.43208) (1024,3.02188) (2048,5.29921) (4096,6.21054) };
    \addlegendentry{algo=Bitonic};
    \addplot coordinates { (0.00411523,1.7899) (0.0123457,1.40477) (0.037037,1) (0.111111,1) (0.333333,1) (1,1) (2,1) (4,1) (8,1.13029) (16,1.75679) (32,2.91817) (64,5.70707) };
    \addlegendentry{algo=FIS};
    \addplot coordinates { (0.00411523,1) (0.0123457,1) (0.037037,1.12762) (0.111111,2.1339) (0.333333,4.46559) (1,9.48848) (2,13.3712) (4,19.7105) (8,29.4864) (16,56.3014) };
    \addlegendentry{algo=Gather};
    \addplot coordinates { (0.00411523,3.3122) (0.0123457,2.77129) (0.037037,2.3426) (0.111111,4.40709) (0.333333,8.9365) (1,39.6267) (2,81.6034) (4,120.219) };
    \addlegendentry{algo=GatherAll};
    \addplot coordinates { (0.00411523,8.8345) (0.0123457,6.89223) (0.037037,4.70873) (0.111111,4.29658) (0.333333,3.61496) (1,2.55642) (2,1.78673) (4,1.30883) (8,1) (16,1) (32,1) (64,1) (128,1) (256,1) (512,1) (1024,1) (2048,1) (4096,1) (8192,1) (16384,1) (33768,1.2634) (65536,1.52284) (131072,1.70695) (262144,2.1046) (524288,2.29799) (1.04858e+06,2.3306) (2.09715e+06,2.40969) (4.1943e+06,2.42044) (8.38861e+06,2.39853) };
    \addlegendentry{algo=Quick};
    \addplot coordinates { (1,26.5796) (2,20.5851) (4,16.6676) (8,12.7971) (16,12.4396) (32,12.0343) (64,11.9848) (128,12.1484) (256,11.4292) (512,10.3878) (1024,5.83753) (2048,5.94952) (4096,3.85926) (8192,2.38707) (16384,1.48761) (33768,1.18885) (65536,1.10038) (131072,1.00168) (262144,1) (524288,1) (1.04858e+06,1) (2.09715e+06,1) (4.1943e+06,1) (8.38861e+06,1) };
    \addlegendentry{algo=ZHYK};
    
    \RunnintimeLegend
\end{axis}
\end{tikzpicture}
\vspace{-0.7cm}
\caption{Uniform}
\end{subfigure}%
\begin{subfigure}[b]{0.5\textwidth}

\begin{tikzpicture}
  \begin{axis}[
    xmax=8388608,
    xmin=0.0041,
    plotconfig,
    ylabel near ticks,
    xlabel near ticks,
    legend style={at={(0,1.25)}, anchor=west},
    legend columns=1, 
    log base x=2,
    xmode=log,
    line width=0.5pt,
    tick style={
      line cap=round,
      thin,
      major tick length=4pt,
      minor tick length=2pt,
    },
  major grid style={thin,dotted,color=black!50},
  minor grid style={thin,dotted,color=black!50},
  grid,
  font=\small,
  ymax=6.5,
  ytick={0, 1, 2, 3, 4, 5, 6},
    ]
    \addplot coordinates { (1,11.7394) (2,8.63328) (4,6.39395) (8,4.65781) (16,4.61605) (32,4.26242) (64,4.01981) (128,3.92348) (256,4.27249) (512,3.89993) (1024,2.08443) (2048,2.46882) (4096,1.6366) (8192,1.22311) (16384,1) (33768,1) (65536,1) (131072,1) (262144,1) (524288,1) (1.04858e+06,1) (2.09715e+06,1) (4.1943e+06,1) (8.38861e+06,1) };
    \addlegendentry{algo=AMS};
    \addplot coordinates { (1,2.65079) (2,1.99243) (4,1.49966) (8,1.15289) (16,1.60839) (32,1.58609) (64,1.69501) (128,1.91988) (256,2.44649) (512,3.44781) (1024,3.01448) (2048,5.37408) (4096,6.42579) };
    \addlegendentry{algo=Bitonic};
    \addplot coordinates { (0.00411523,1.7932) (0.0123457,1.39835) (0.037037,1) (0.111111,1) (0.333333,1) (1,1) (2,1) (4,1) (8,1.08115) (16,1.72298) (32,2.81542) (64,5.55746) };
    \addlegendentry{algo=FIS};
    \addplot coordinates { (0.00411523,1) (0.0123457,1) (0.037037,1.11565) (0.111111,2.15124) (0.333333,4.45859) (1,9.63964) (2,13.9263) (4,19.2425) (8,28.685) (16,57.0672) };
    \addlegendentry{algo=Gather};
    \addplot coordinates { (0.00411523,3.32181) (0.0123457,2.77645) (0.037037,2.32511) (0.111111,4.46265) (0.333333,8.97535) (1,40.2793) (2,83.7666) (4,124.438) };
    \addlegendentry{algo=GatherAll};
    \addplot coordinates { (0.00411523,8.88553) (0.0123457,6.84174) (0.037037,4.70266) (0.111111,4.43661) (0.333333,3.59135) (1,2.59312) (2,1.82937) (4,1.36577) (8,1) (16,1) (32,1) (64,1) (128,1) (256,1) (512,1) (1024,1) (2048,1) (4096,1) (8192,1) (16384,1.09724) (33768,1.36338) (65536,1.54345) (131072,1.63184) (262144,2.45052) (524288,2.08314) (1.04858e+06,2.48072) (2.09715e+06,2.61897) (4.1943e+06,2.67744) (8.38861e+06,2.70296) };
    \addlegendentry{algo=Quick};

    \legend{}
  \end{axis}
\end{tikzpicture}
\vspace{-0.3cm}
\caption{BucketSorted}
\end{subfigure}
\\\vspace{0.3cm}
\begin{subfigure}[b]{0.5\textwidth}

\begin{tikzpicture}
  \begin{axis}[
    xmax=8388608,
    xmin=0.0041,
    plotconfig,
    ylabel near ticks,
    xlabel near ticks,
    ylabel={Running Time Ratio $t/t_{\text{best}}$},
    legend columns=1, 
    log base x=2,
    xmode=log,
    line width=0.5pt,
    tick style={
      line cap=round,
      thin,
      major tick length=4pt,
      minor tick length=2pt,
    },
  major grid style={thin,dotted,color=black!50},
  minor grid style={thin,dotted,color=black!50},
  grid,
  xlabel={$\frac{n}{p}$},
  font=\small,
  legend pos=north east,
  legend style={font=\normalsize},
  ymax=6.5,
  ytick={0, 1, 2, 3, 4, 5, 6},
    ]
    \addplot coordinates { (1,11.8104) (2,8.70645) (4,6.41121) (8,4.68076) (16,4.48029) (32,4.25555) (64,4.00835) (128,4.3302) (256,4.13812) (512,3.74435) (1024,2.24612) (2048,2.40527) (4096,1.79765) (8192,1.39617) (16384,1) (33768,1) (65536,1) (131072,1) (262144,1) (524288,1) (1.04858e+06,1) (2.09715e+06,1) (4.1943e+06,1) (8.38861e+06,1) };
    \addlegendentry{algo=AMS};
    \addplot coordinates { (1,2.68804) (2,2.00427) (4,1.48994) (8,1.15265) (16,1.55837) (32,1.57642) (64,1.66752) (128,1.91694) (256,2.42322) (512,3.40787) (1024,2.99784) (2048,5.2124) (4096,6.10152) };
    \addlegendentry{algo=Bitonic};
    \addplot coordinates { (0.00411523,1.91989) (0.0123457,1.52322) (0.037037,1) (0.111111,1) (0.333333,1) (1,1) (2,1) (4,1) (8,1.08188) (16,1.71573) (32,2.851) (64,5.61551) };
    \addlegendentry{algo=FIS};
    \addplot coordinates { (0.00411523,1) (0.0123457,1) (0.037037,1.00432) (0.111111,1.97549) (0.333333,4.04625) (1,8.53718) (2,11.3582) (4,16.6475) (8,22.78) (16,42.6434) };
    \addlegendentry{algo=Gather};
    \addplot coordinates { (0.00411523,3.35586) (0.0123457,2.8709) (0.037037,2.23852) (0.111111,4.54323) (0.333333,9.19612) (1,40.7701) (2,84.7374) (4,122.941) };
    \addlegendentry{algo=GatherAll};
    \addplot coordinates { (0.00411523,8.95531) (0.0123457,7.13933) (0.037037,4.53728) (0.111111,4.39984) (0.333333,3.68281) (1,2.62136) (2,1.85802) (4,1.36916) (8,1) (16,1) (32,1) (64,1) (128,1) (256,1) (512,1) (1024,1) (2048,1) (4096,1) (8192,1) (16384,1.02035) (33768,1.25378) (65536,1.3816) (131072,1.37886) (262144,1.53226) (524288,1.58606) (1.04858e+06,1.46373) (2.09715e+06,1.38316) (4.1943e+06,1.31772) (8.38861e+06,1.27236) };
    \addlegendentry{algo=Quick};
    \addplot coordinates { (1,25.052) (2,19.4441) (4,15.6032) (8,11.3572) (16,11.1178) (32,10.7128) (64,10.6183) (128,10.9066) (256,10.1976) (512,9.05067) (1024,5.07426) (2048,5.21276) (4096,3.76696) (8192,2.70978) (16384,2.00125) (33768,1.8759) (65536,1.88392) (131072,1.69457) (262144,1.68715) (524288,1.64283) (1.04858e+06,1.50508) (2.09715e+06,1.36739) (4.1943e+06,1.2845) (8.38861e+06,1.24241) };
    \addlegendentry{algo=ZHYK};
    
    \legend{}
  \end{axis}
\end{tikzpicture}
\vspace{-0.3cm}
\caption{Staggered}
\end{subfigure}%
\begin{subfigure}[b]{0.5\textwidth}

\begin{tikzpicture}
  \begin{axis}[
    xmax=8388608,
    xmin=0.0041,
    plotconfig,
    xlabel near ticks,
    legend columns=1, 
    log base x=2,
    xmode=log,
    line width=0.5pt,
    tick style={
      line cap=round,
      thin,
      major tick length=4pt,
      minor tick length=2pt,
    },
  major grid style={thin,dotted,color=black!50},
  minor grid style={thin,dotted,color=black!50},
  grid,
  xlabel={$\frac{n}{p}$},
  ymax=6.5,
  ytick={0, 1, 2, 3, 4, 5, 6},
  font=\small,
  compat=1.3,
  y label style={at={(-0.08,0.5)}},]
    \addplot coordinates { (1,11.896) (2,8.59496) (4,6.3845) (8,4.53321) (16,4.40312) (32,4.14312) (64,3.94838) (128,3.91552) (256,4.33285) (512,2.75054) (1024,2.16752) (2048,2.10596) (4096,1.77117) (8192,1.22621) (16384,1) (33768,1) (65536,1) (131072,1) (262144,1) (524288,1) (1.04858e+06,1) (2.09715e+06,1) (4.1943e+06,1) (8.38861e+06,1) };
    \addlegendentry{algo=AMS};
    \addplot coordinates { (1,2.72023) (2,2.00391) (4,1.49462) (8,1.1212) (16,1.52961) (32,1.55058) (64,1.66932) (128,1.89568) (256,2.38321) (512,2.35917) (1024,2.96975) (2048,4.87565) (4096,6.11369) };
    \addlegendentry{algo=Bitonic};
    \addplot coordinates { (0.00411523,2.02706) (0.0123457,1.64143) (0.037037,1.14789) (0.111111,1) (0.333333,1) (1,1) (2,1) (4,1) (8,1.0414) (16,1.66181) (32,2.79048) (64,5.52869) };
    \addlegendentry{algo=FIS};
    \addplot coordinates { (0.00411523,1) (0.0123457,1) (0.037037,1) (0.111111,1.49436) (0.333333,2.87587) (1,5.94966) (2,8.41529) (4,12.5731) (8,17.4222) (16,33.8428) };
    \addlegendentry{algo=Gather};
    \addplot coordinates { (0.00411523,3.85231) (0.0123457,3.44078) (0.037037,2.6277) (0.111111,4.14589) (0.333333,8.11306) (1,38.7142) (2,80.935) (4,120.223) };
    \addlegendentry{algo=GatherAll};
    \addplot coordinates { (0.00411523,9.42334) (0.0123457,7.70506) (0.037037,5.42654) (0.111111,4.41144) (0.333333,3.61654) (1,2.64237) (2,1.84841) (4,1.37809) (8,1) (16,1) (32,1) (64,1) (128,1) (256,1) (512,1) (1024,1) (2048,1) (4096,1) (8192,1) (16384,1.05502) (33768,1.15198) (65536,1.2831) (131072,1.34927) (262144,1.53318) (524288,1.59987) (1.04858e+06,1.58964) (2.09715e+06,1.62707) (4.1943e+06,1.60857) (8.38861e+06,1.57253) };
    \addlegendentry{algo=Quick};
    
    \legend{}
\end{axis}
\end{tikzpicture}
\vspace{-0.3cm}
\caption{DeterDupl}
\end{subfigure}
\caption{Running  ratios of each algorithm to the fastest algorithm on 262\,144 cores. \ahyk crashes on input instances \deterdupl and \bucketsorted.}
\label{fig:largest instance ratio}
\end{figure*}
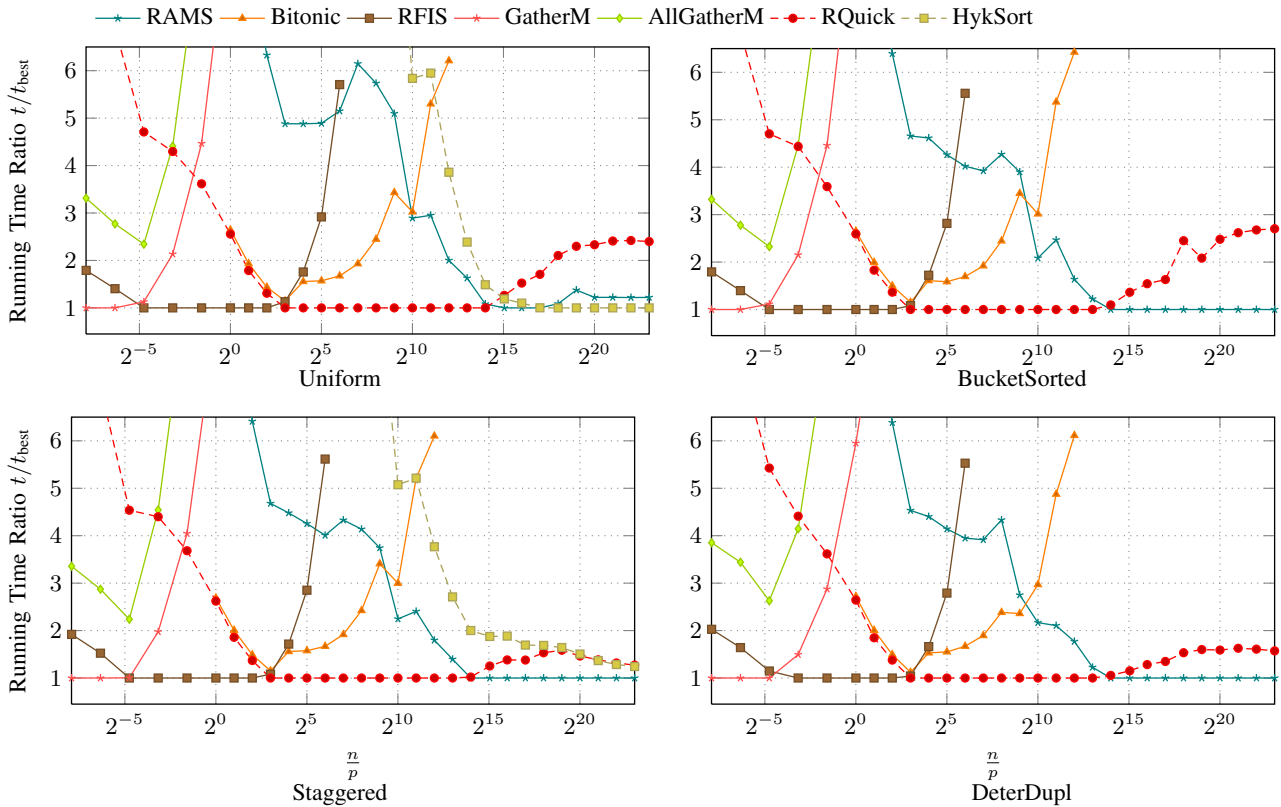

We call inputs \emph{dense} when $n/p \geq 1$ and sparse when $n/p <1$. \emph{Sparsity-factor} $i$ means that only every $i$-th PE holds an input elements.
The implementations \agather, \aallgather, \afis, \aquick, and \aams are in general capable of sorting all inputs.
Exceptions are \abitonic which fails to sort sparse inputs and \ahyk which fails to sort inputs with many duplicates.

For \afis, we proposed in Subsection~\ref{s:fastInefficient ams} a hypercube exchange operation to route the elements to their target PEs.
In contrast, our implementation delivers elements in single messages which is more efficient for $n/p < \log p$.
For those input sizes, we expect \afis to be faster than its competitors.
Axtmann el al.~\cite{Axtmann15} use a scattered all-reduce operation to reduce the local ranks in their variant of FIS.
We use \texttt{MPI\_Allreduce} instead without observing significant overheads due to this decision.

$l$-level \aams selects $b = \frac{2}{\sqrt[l]{1 + \epsilon} - 1}$ splitters to guarantee a maximum imbalance of $\epsilon=0.2$.
In our experiments, the imbalance was always smaller than $0.1$ (except for $n/p \leq 16$).
\asamplesort uses a sample size of $16 \log p$ elements per PE and calls \texttt{MPI\_Alltoallv} to exchange data. The implementation uses the \texttt{MPI\_Datatype} \texttt{MPI\_Byte} for every \verb!C++! data type.
When we used other \texttt{MPI\_Datatypes}, the \texttt{MPI\_Alltoallv} call ran into a machine specific deadlock.
Sundar et al.~\cite{Sundar13} implemented a nonrobust and a robust version of $k$-way sample sort which both use \texttt{MPI\_Alltoall} and \texttt{MPI\_Alltoallv}.
We did not include their implementation in our benchmarks as these MPI-calls deadlocked.
The deadlocks may also be a machine specific problem as the implementation successfully ran on our institute server.
Siebert and Wolf \cite{SieWol11} developed \emph{\aminisort} for the case $n = p$. Unfortunately, the exact source code used in their experiments is not available any more.
The algorithms are written in \verb!C++11! and compiled with version~$4.8.1$ of
the GNU compiler collection, using the optimization flag \emph{-O2}. For inter-process
communication, we used the IBM \verb!mpich2! library version 1.5, Argonne.

\subsubsection{Parameter Tuning}
\label{app:parameter tuning}
We performed extensive parameter tuning on 131\,072 cores, half of the available number due to a limited number of available core hours.
Each algorithm has a range of input sizes where it is designed to be the fastest.
We optimized each algorithm to run as fast as possible in this input range.
Sundar et al.~\cite{Sundar13} parallelized the initial sorting step of \abitonic.
In our experiments, we use \abitonic with one thread per process since the merging step is not parallelized.
Just for $n=p$, using more threads per process is slightly faster than \abitonic with one thread per process.
We tested \ahyk with all combinations of $k=16, 32, 64$ and $2, 8, 16$ threads per process and sorted \uniform inputs of up to $2^{20}$ elements.
\ahyk did not terminate when we executed the algorithm with one thread per process.
\ahyk with $k=32$ and $8$ threads is the fastest algorithm overall (or is as fast as \aams) for at least $2^{17}$ elements.
\aams with three levels is the fastest algorithm overall for input of size $2^{15}$ and $2^{16}$.
More levels speed up \aams for small inputs (up to $50\%$) and less levels slightly speed up \aams for larger inputs.

\subsection{Algorithm Comparison with Running Time Rations}
\label{app:largest instance ratio}

Figure~\ref{fig:largest instance ratio} shows the running time ratios of each algorithm compared to the fastest algorithm for the most interesting input instances.

\subsection{Comparison to the Literature}
\label{app:comparison to the literature}

We compare our results to the literature and the web. MP-sort \cite{FSSMC14} is a single-level multiway mergesort
running on up to 160\,000 cores of a Cray XE6.
For 128 GB of data it is 3\,635 times slower than ours for $p=262\,144$ and 137 GB.
The 2015 winner of the Sort Benchmark (\url{sortbenchmark.org}), FuxiSort~\cite{fixi15, zhang2014fuxi} sorts 7.7 TB of 100 byte elements with 10 byte random keys using 41\,496 cores in 58s (MinuteSort).  That is 9 times longer than \aams and \ahyk take for sorting 1.1 times as much data and 15.7 times more (8-byte) elements
using 6.3 times more cores. FuxiSort has to read/write from/to disks. However, the disks of that system are not the bottleneck.

\subsection{Acknowledgment}
\label{app:ack}

The authors gratefully acknowledge the Gauss Centre for Supercomputing (GCS) for providing computing time through the John von Neumann Institute for Computing (NIC) on the GCS share of the supercomputer JUQUEEN~\cite{stephan2015juqueen} at J{\"u}lich Supercomputing Centre (JSC). GCS is the alliance of the three national supercomputing centres HLRS (Universität Stuttgart), JSC (Forschungszentrum J{\"u}lich), and LRZ (Bayerische Akademie der Wissenschaften), funded by the German Federal Ministry of Education and Research (BMBF) and the German State Ministries for Research of Baden-W{\"u}rttemberg (MWK), Bayern (StMWFK) and Nordrhein-Westfalen (MIWF).
Special thanks go to SAP AG, Ingo M\"uller, and Sebastian Schlag for making their 1-factor
algorithm~\cite{SSM13} available.
We gratefully thank Helman et al.~\cite{Helman98} which provided generators for seven input instances and Sundar et al.~\cite{Sundar13} for making HykSort available.
This research was partially supported by DFG
project SA 933/11-1.
\end{document}